\documentclass[aps,twocolumn,amsmath,amssymb,preprintnumbers]{revtex4}
\usepackage{amsmath} \usepackage{amsfonts} \usepackage{amssymb}
\usepackage{bbm}
\usepackage{epsfig}
\usepackage{graphics}
\usepackage{graphicx}
\newcommand{\be}{\begin{equation}}
\newcommand{\ee}{\end{equation}}
\newcommand{\ba}{\begin{eqnarray}}
\newcommand{\ea}{\end{eqnarray}}
\newcommand{\nn}{\nonumber}
\newcommand{\kr}{\rangle}
\newcommand{\kl}{\langle}

\newcommand{\tr}{\textup{tr}}

\begin{document}

\title[ ]{Quantum particles from classical probabilities in phase space}

\author{C. Wetterich}
\affiliation{Institut  f\"ur Theoretische Physik\\
Universit\"at Heidelberg\\
Philosophenweg 16, D-69120 Heidelberg}

\begin{abstract}
Quantum particles in a potential are described by classical statistical probabilities. We formulate a basic time evolution law for the probability distribution of classical position and momentum such that all known quantum phenomena follow, including interference or tunneling. The appropriate quantum observables for position and momentum contain a statistical part which reflects the roughness of the probability distribution. ``Zwitters'' realize a continuous interpolation between quantum and classical particles. Such objects may provide for an effective one-particle description of classical or quantum collective states as droplets of a liquid, macromolecules or a Bose-Einstein condensate. They may also be used for quantitative fundamental tests of quantum mechanics. 
We show that the ground state for zwitters has no longer a sharp energy. This feature permits to put quantitative experimental bounds on a small parameter for possible deviations from quantum mechanics. 
\end{abstract}

\maketitle

\section{Introduction}
\label{Intro}

An effective one-particle picture may be useful for collective systems in many circumstances. An example is a droplet of water. For many problems we may wish to reduce its description to states characterized only by position and momentum variables $x$ and $p$. In many circumstances such a description will be probabilistic, based on a probability distribution in phase space $w(x,p)$. A problem can find an effective one-particle description if the time evolution of $w(x,p)$ can be computed from the knowledge of $w(x,p)$ alone, typically in the form of a differential equation 
\be\label{N1}
\partial_t w(x,p;t)=F\big[w(x,p;t)\big],
\ee
with $F[w]$ a functional of $w$. This should hold at least approximately. In case of an evolution equation \eqref{N1} the information contained in $w(x,p;t_0)$ at some initial time $t_0$ is sufficient in order to predict the dynamics at later times.

The time evolution of the probability distribution for the particle will depend on its environment. We assume here a stationary situation for the environment for which the space dependence can be encoded in a function $V(x)$. A classical particle would feel an effective force given by the gradient of a potential $V(x)$, but we will consider here more general situations how the environment influences the dynamics. Interesting experiments with droplets of a liquid have indeed created an environment where the observed motion is quite different from classical particles \cite{Cov}. In analogy to the potential we will assume that $V(x)$ has the dimension of an energy, while the interpretation may differ from a static potential energy. The aim of this paper is an investigation how different possible forms of the evolution equation \eqref{N1} influence the dynamics of the ``particle''. We find a particular form of this evolution equation for which eq. \eqref{N1} becomes effectively the Schr\"odinger equation for a quantum particle in a potential $V(x)$. 

For a classical point particle the evolution equation \eqref{N1} is given by the Liouville equation,
\ba\label{N2}
\partial_tw&=&-\hat L w,\nn\\
\hat L&=&\frac{p}{m}\partial_x-\frac{\partial V}{\partial x}\partial_p
\ea
with $m$ the mass of the particle and $V(x)$ some external potential. Here $x,p$ are the position and momentum in six-dimensional phase space, with easy generalization to a lower dimensionality of space. In general, the dynamics of droplets will differ from classical point particles, and one expects an effective evolution equation different from eq. \eqref{N2}. Indeed, the droplet has many internal degrees of freedom or microstates that have to be integrated out in order to arrive at a probability distribution in phase space and its time evolution \eqref{N1}. One obtains $w(x,p)$ by selecting some appropriate observables $x$ and $p$ and summing over the probabilities of all microstates of the droplet that have common values of $x$ and $p$. The effective evolution equation obtains by averaging over the detailed dynamics of the microstates. This is, in general, a complicated procedure. The precise form of eq. \eqref{N1}, including the possibility to find such a simplification at all, depends on the choice of $x$ and $p$. In this paper we will not attempt to compute the evolution equation \eqref{N1} from microphysics. We will rather explore some general consequences of evolution equations that differ from the Liouville equation \eqref{N2}. Besides classical systems as droplets or other collective systems with internal degrees of freedom one may also envisage an effective one-particle description of collective quantum systems as macromolecules or a Bose-Einstein condensate. 

One may even take the point of view that for microscopic particles a probabilistic description is always appropriate, and that the basic evolution law \eqref{N1} is not known a priori. The Liouville equation for point particles may emerge only in some suitable ``classical limit''. We will see that on a more fundamental level the generalized setting of eq. \eqref{N1} allows for the description of a microscopic quantum particle in a potential, provided we choose an appropriate form of the evolution equation. In this case, the Schr\"odinger equation will be derived from a coarse graining of the evolution equation \eqref{N1}.

Indeed, it has been shown \cite{3A,CWAA1} that quantum particles and classical particles can be described within the same probabilistic framework. For both one can use as a basic concept a probability distribution $w(x,p)$ which depends on position $x$ and momentum $p$. The associated real wave function in phase space, $\psi_C(x,p)$, obeys $w(x,p)=\psi^2_C(x,p)$. This common description shows new features both for the classical particle and for the quantum particle. For the classical particle we abandon the trajectories as a basic concept - sharp trajectories follow only as a limiting case for infinitely sharp probability distributions $w(x,p)$. The particle - wave duality, often believed to be characteristic for quantum physics, also applies to classical particles. For a quantum particle a real wave function depending on {\em both} momentum {\em and} position is unusual. We will see how $\psi_C(x,p)$ can be related to the familiar complex quantum wave function $\psi_Q(x)$ or $\psi_Q(p)$ which depends {\em either} on position {\em or} on momentum, but not on both.

We will formulate the basic law for the dynamics of a particle as an evolution law for the classical wave function. We will write it as a first order differential equation in time for $\psi_C(x,p)$, in analogy to the Schr\"odinger equation for quantum mechanics. Such an evolution equation can be directly related to the associated evolution equation \eqref{N1} for the probability density.  For a classical particle, the evolution equation for the classical wave function is equivalent to the Liouville equation, while for a quantum particle $\psi_C(x,p)$ follows a different evolution law. The latter will be shown to be equivalent to the Schr\"odinger equation for the quantum wave function $\psi_Q(x)$, or the associated von-Neumann equation for the density matrix. 

A one-particle description of water droplets or Bose-Einstein condensates will neither behave exactly as a classical nor as a quantum particle. The effective evolution equation for such ``particles'' will be somewhere inbetween, with detailed form depending on the concrete collective system. In order to get a first glance on the possibilities we investigate here simple interpolations between the quantum and classical behavior. Indeed, since standard classical particles and quantum particles can be described within the same setting, only differing by the time evolution of the probability distribution, we can consider the possibility of ``zwitters'' - particles with properties interpolating between the quantum particle and the classical particle as a function of some continuous parameter $\gamma$ \cite{3A}. While for $\gamma=0$ a quantum particle shows interference in a double slit experiment, the classical particle for $\gamma=\pi/2$ passes through only one of the slits without interference effects. Particularly interesting systems are close to quantum particles with a small parameter $\gamma$, since they permit to quantify fundamental tests of the validity of quantum mechanics. Precision measurements can test the ``quantumness'' in a quantitative way by putting bounds on $\gamma$.

One may also view the formalism developed in the present paper as a realization of quantum mechanics within the framework of classical statistics \cite{3A,CWAA1}. The material presented in this work therefore serves a double purpose. On the one side we attempt first steps for the understanding of generalized evolution equations for an effective one-particle description of collective systems. On the other side it may be considered as a contribution to the interpretation of quantum mechanics. In this context we emphasize that our implementation of a quantum particle within classical statistics is not a deterministic local hidden variable theory. It rather assumes that the fundamental description of the real world is intrinsically probabilistic. In ref. \cite{CWAA1,CW2,CWE} the basic conceptual settings how quantum physics can emerge from a classical statistical ensemble have been described in detail. It was shown how the quantum formalism with non-commuting operators arises form a description of a subsystem. We have discussed explicitly quantum mechanically entangled states \cite{CW1}, \cite{CW2}, \cite{CWE} and shown that the measurement correlation $\kl AB\kr_m$ is equivalent to the usual quantum correlation and violates Bell's inequalities \cite{Bell}.

In the present paper we present an explicit non-linear time evolution equation for the phase space density $w(x,p)$ such that all expectation values and correlations of position and momentum observables take the same values as for a quantum particle in a potential. This clearly demonstrates that a quantum particle can be described by an appropriate classical statistical ensemble, including the striking phenomena of interference and tunneling. The exact coincidence of all measurable quantities with the predictions of quantum mechanics holds provided we choose the appropriate observables for position and momentum and the appropriate measurement correlation based on conditional probabilities.

We have shown in ref. \cite{CWAA} that it is possible, in principle, to formulate a classical statistical ensemble for a quantum particle such that the position and momentum observables can be associated to standard classical observables which take a fixed value in every state of some microscopic ensemble. However, for the description of a quantum particle by a probability distribution in phase space $w(x,p)$ it is not clear a priori if the position and momentum observables should be directly associated with $x$ and $p$. We will actually see that this is not the case. 

In the present paper we include in our statistical formulation a new type of ``statistical observables'' \cite{CWAA} that measure statistical properties of the probability distribution as its roughness in position or momentum. The expectation value of such a statistical observable is computable for any given $w(x,p)$. However, statistical observables do not assume fixed values for a given point in phase space $(x,p)$. (In this respect, statistical observables are conceptually similar to entropy in a thermodynamic equilibrium ensemble. In contrast to entropy, they are defined for every given $w(x,p)$ and do not require a particular equilibrium configuration.) By the use of statistical observables the probabilistic description of a quantum particle gets rather simple and can be achieved in terms of the probability density in phase space $w(x,p)$. The statistical observables may ultimately be derived from standard classical observables formulated on a more microscopic level, but this is not essential in our context. 

For the possible continuous interpolations between a quantum particle and a classical particle we discuss both the possibility of sharpened observables for quantum particles and the effects of a time evolution interpolating between quantum and classical particles. We denote by ``zwitters'' those effective one-particle systems whose dynamics is described by an interpolating time evolution. They behave neither purely quantum nor purely classical. In a larger sense, a water droplet or a Bose-Einstein condensate whose dynamics is characterized by an effective evolution equation \eqref{N1} may be called a zwitter. Even though the definition of the interpolation is not unique, it is instructive to discuss zwitters for a specific realization of the interpolation. The specific interpolation discussed in the present paper results in a broadening of the energy level of the ground state for zwitters, as compared to a sharp ground state energy for a quantum particle. Measuring the energy width of the ground state can therefore experimentally constrain a small parameter $\gamma$ for deviations from quantum mechanics. 

The very possibility of a continuous interpolation between quantum and classical particles within a formulation based on a classical statistical  probability distribution in phase-space is perhaps the most convincing demonstration that there is no conceptual jump between quantum and classical. We emphasize that this is not the same as taking the classical limit in the quantum formalism. The latter can be done easily for the Wigner function by letting $\hbar\to 0$. However, the Wigner function is not a probability density.

This paper is organized as follows: In sect. \ref{Classical wave function} we recall the probabilistic formulation for a single classical particle in terms of a wave function \cite{CWAA}. The concept of the classical wave function allows for the introduction of statistical observables which can be computed from the probability density for a classical particle. They are intimately related to the basic probabilistic formulation and find no appropriate role for single classical trajectories. In sect. \ref{Quantumparticlesfrom} we describe the quantum particle in the same formulation as the classical particle. However, the choice of observables for position and momentum and the basic time evolution law for the probability density are now modified. We discuss in sect. \ref{Correlations and incomplete statistics} several conceptual issues in quantum mechanics that find an analogue in the coarse graining of the information that occurs in the translation from a statistical ensemble of molecules to a statistical description of a droplet.

Sect. \ref{Quantum density matrix} addresses the quantum density matrix that can be associated to any given classical wave function $\psi_C(x,p)$. This establishes how the usual quantum formalism with pure state wave functions either depending on position or on momentum, but not on both, is embedded in our description of a wave function in phase space.

In sect. \ref{Quantum and classical energy} we turn to the concept of energy which shows important differences between the classical and quantum objects. Sect. \ref{Continuousinterpolation} discusses possible continuous interpolations between a classical particle and a quantum particle. In sect. \ref{Sharpened observables} the interpolation of observables is investigated, while sect. \ref{Zwitters - continuous} explores the interpolation of the evolution equation and addresses experimental tests for ``particle models'' in the vicinity of quantum mechanics. Our conclusions are presented in sect. \ref{conclusionsand}.

\section{Classical wave function and statistical observables}
\label{Classical wave function}

The basic concept of this section is the ``classical'' wave function in phase-space $\psi_C(x,p)$. This real function is related to the probability density in phase-space, $w(x,p)$, by
\ba\label{85B}
w(x,p)=\psi^2_C(x,p).
\ea
Up to a sign the classical wave function is fixed by $w(x,p)$
\be\label{Q1}
\psi_C(x,p)=s(x,p)\sqrt{w(x,p)}~,~s(x,p)=\pm 1.
\ee
As we have discussed in detail in ref. \cite{CWAA}, this sign is largely determined by analyticity properties. The classical wave function plays for the description of a single classical particle the same role as the quantum wave function for a quantum particle. It is a normalized probability amplitude
\be\label{Q2}
\int_{x,p}\psi^2_C(x,p)=1,
\ee
where $\int_x=\int d^3x,\int_p=\int d^3p/(2\pi)^3$ and we use $\hbar=1$ unless stated otherwise. In contrast to the quantum wave function, $\psi_C(x,p)$ depends both on $x$ and $p$ and is real. The definition \eqref{85B} guarantees that $w$ is positive, $w(x,p)\geq 0$, and the normalization \eqref{Q2} ensures the normalization of the probability density. 

The expectation values of observables are computed according to the usual quantum rule
\be\label{Q3}
\kl A\kr=\int_{x,p}\psi_C(x,p)A\psi_C(x,p).
\ee
In particular, the classical position and momentum observables $X_{cl}$ and $P_{cl}$ commute and can be represented by $x$ and $p$. Functions of classical position and momentum obey 
\be\label{85A}
\kl F(\hat X_{cl},\hat P_{cl})\kr=
\int_{x,p}\psi^2_C(x,p)F(x,p),
\ee
in accordance with the classical statistical rule in terms of $w(x,p)$. 

The classical wave function $\psi_C(x,p)$ is the central object for casting a probabilistic theory of classical particles into the quantum formalism. Particle-wave duality applies to a classical particle just as for a quantum particle. While the detection of a particle is a discrete yes/no event of finding the particle in a detector or not, the probability amplitude $\psi_C(x,p)$ carries the continuous wave aspects.

The notion of trajectories is not basic in our probabilistic description - classical trajectories only arise as limiting cases if $w(x,p)$ is an arbitrarily sharp distribution. In consequence, we cannot use Newton's laws for trajectories as a basic formulation of the dynamics of the theory. The basic dynamic equation should rather be formulated as a time evolution equation for the probability density $w(x,p)$. We will formulate it as a linear equation for the classical wave function 
\be\label{89A}
i\partial_t\psi_C(x,p)=H\psi_C(x,p).
\ee
Here we have used the form of a Schr\"odinger equation in order to make the analogy to the quantum formulation apparent. Since $\psi_C(x,p)$ is a real function, $H$ should be a purely imaginary operator. We stress that the evolution equation for $\psi_C$ defines the evolution equation for $w$ in an unambiguous way since $w=\psi^2_C$. While eq. \eqref{89A} is linear in $\psi_C$, the equivalent equation for $w(x,p$ may become non-linear. 

For an appropriate choice of $H$ the time evolution of the probability distribution is determined by the Liouville equation, as standard for a classical particle. This is the case if $H$ equals the Liouville operator up to a factor $i$
\be\label{Q4}
H_L=-i\hat L=-i\frac pm\partial_x+i\frac{\partial V}{\partial x}\partial_p.
\ee
Since $H_L$ is linear in the derivatives $\partial_x$ and $\partial_p$ the evolution equation for $w$ remains linear as well in this case,
\be\label{Q5}
\partial_t w=-\hat L w.
\ee
It is well known that the Liouville equation \eqref{Q5} follows from Newton's laws for statistical ensembles of classical particles. We employ it here as the basic dynamical equation for a single classical particle. In turn, Newton's laws for trajectories of classical particles follow from eq. \eqref{Q5} in the limit of infinitely sharp probability distributions. For any nonzero width of $w(x,p)$, however, the distribution will broaden in the course of the time evolution \cite{CWAA,Vol}, as well known from the dispersion of quantum particles. 

With the use of the Liouville operator \eqref{Q4} our discussion of the classical wave function shares some important features with the Hilbert space formulation of classical mechanics by Koopman and von Neumann \cite{Koop}. Linear evolution equations of this type have been studied for complex wave functions depending on $x$ and $p$ \cite{Ma}. Furthermore, the Hilbert space formulation of classical mechanics has triggered many interesting formal developments \cite{GR}. We emphasize, however, that for our approach it is crucial that $\psi_C(x,p)$ is a {\em real} function. A complex wave function contains an additional degree of freedom associated to its phase and cannot be constructed from $w(x,p)$ (up to irrelevant sign ambiguities). We also will modify the time evolution by choosing $H$ different from $H_L$. (In a different context and conceptual setting a nonlinear evolution of a wave function for classical particles has been discussed in ref. \cite{NLW}.)

The concept of a classical wave function with expectation values of observables defined by eq. \eqref{Q3} permits the use of ``derivative observables'' that are not familiar in classical statistics. For example, we may use the operators
\be\label{Q6}
X_s=i\partial_p~,~P_s=-i\partial_x.
\ee
Even powers of $X_s$ or $P_s$ are real operators and can be interpreted as observables in the quantum formalism. (The expectation values of odd powers vanish.) For example, the expectation value of $P^2_s$ is defined by eq. \eqref{Q3}
\be\label{Q7}
\kl P^2_s\kr=\int_{x,p}\big(\partial_x\psi_C(x,p)\big)^2,
\ee
which differs from $-\int_{x,p}\partial^2_xw(x,p)=0$. One finds that $\kl P^2_s\kr$ measures the ``roughness''  of the probability distribution in position space \cite{CWAA}. We can compute $\kl P^2_s\kr$ in terms of the probability density $w$ without using $\psi$, but the expression is non-linear
\be\label{Q8}
\kl P^2_s\kr=\frac14\int_{x,p}w^{-1}(\partial_xw)^2=\frac14\int_{x,p}w
(\partial_x\ln w)^2,
\ee
where we use $\partial_xw=2\psi_C\partial_x\psi_C$. The use of the ``statistical observables'' $X_s$ and $P_s$ will become apparent in the next section where we describe a quantum particle in terms of the classical probability distribution $w(x,p)$.

\section{Quantum particles from classical probabilities}
\label{Quantumparticlesfrom}

In this section we show explicitly how all the properties of a quantum particle can be described in terms of classical probabilities. For this purpose we use the classical wave function $\psi_C(x,p)$, which encodes the information about the classical probability distribution $w(x,p)=\psi^2_C(x,p)$. As compared to the classical particle, the quantum particle needs two modifications: (i) the use of quantum observables instead of classical observables, and (ii) a different time evolution of the classical probability density in phase space.

\medskip\noindent
{\bf 1. \quad Quantum observables}

The quantum observables $X_Q$ and $P_Q$ for position and momentum are defined by the operators 
\be\label{F12}
X_Q=x+\frac i2\frac{\partial}{\partial p}~,~P_Q=p-\frac i2\frac{\partial}{\partial x},
\ee
They obey the usual commutation relation in quantum mechanics
\be\label{F13}
[X_Q,P_Q]=i.
\ee
The expectation value for an arbitrary sequence of such observables is computed from the classical probability distribution by the usual quantum rule
\be\label{J1}
\kl F(X_Q,P_Q)\kr=\int_{x,p}\psi_C(x,p)F(X_Q,P_Q)\psi_C(x,p).
\ee
We observe that the quantum observables involve the statistical observables $X_s$ and $P_s$. 
\be\label{Q9}
X_Q=X_{cl}+\frac12 X_s~,~P_s=P_{cl}+\frac12 P_s.
\ee

A convenient tool  for the computation of expectation values of products of $X_Q$ and $P_Q$ is the Wigner function $\rho_w(x,p)$ \cite{Wig}, \cite{Moyal}. It is defined here in terms of the classical wave function $\psi_C(x,p)$ by
\ba\label{AA1}
&&\hspace{-0.2cm}\rho_w(x,p)=\\
&&\hspace{-0.2cm}\int_{r,r',s,s'}
\psi_C(x+\frac r2,p+s)\psi_C(x+\frac{r'}{2},p+s')\cos (s'r-sr').\nn
\ea  
Since the classical wave function $\psi_C$ is directly related to the probability distribution $w$ by eq. \eqref{Q1}, this Wigner function can be expressed in terms of the probability density in phase space. We may call the map from $w(x,p)$ to $\rho_w(x,p)$ the quantum transform of $w$. (It requires an appropriate choice of the sign function $s(x,p)$.) We also emphasize that at this point the quantum wave function or density matrix is not yet introduced for the quantum particle. The choice of the quantum transform \eqref{AA1} can be motivated by a coarse graining procedure that we have discussed more extensively ref. \cite{3A}. One can verify by explicit calculation that the expectation values of totally symmetrized products of $X_Q$ and $P_Q$ obey the simple relation
\be\label{Q10}
\kl F_s(X_Q,P_Q)\kr=\int_{x,p}F(x,p)\rho_w(x,p).
\ee
This is the same expression for expectation values in terms of the Wigner function as in quantum mechanics. We can therefore identify the Wigner function \eqref{AA1} with the standard Wigner function of a quantum particle. Since $X_Q$ and $P_Q$ have also the same commutator as in quantum mechanics, all correlation functions of $X_Q$ and $P_Q$ which are computed by the rule \eqref{Q3} from the probability density in phase space $w(x,p)$ are precisely the same as the ones computed for a quantum particle with a quantum density matrix or wave function that leads to the same Wigner function $\rho_w(x,p)$. 

We emphasize that the concepts of the classical wave function, the statistical observables, the Wigner function \eqref{AA1} and the quantum observables \eqref{F12} can be defined for an arbitrary one-particle probability distribution $w(x,p)$. They do not depend on the issue if the time evolution is given by the classical evolution or something else. This recalls us that for collective systems as water droplets some thought is needed for a determination of a position and momentum observables that correspond best to a given type of measurement.

\medskip\noindent
{\bf 2. \quad Quantum evolution of probabilities in phase

{~}\hspace{0.3cm}  space}

If we keep the classical dynamical law for the evolution of the probability density in terms of the Liouville equation \eqref{89A}, \eqref{Q4}, \eqref{Q5}, the time evolution of the correlation functions for $X_Q$ and $P_Q$ will differ from the time evolution in quantum mechanics. The classical time evolution does not induce the standard von Neumann equation for $\rho_w$. Actually, the quantum correlations defined by products of $X_Q$ and $P_Q$ are interesting objects even for a classical particle governed by the Liouville equation. They constitute a possible alternative for the observables measuring position and momentum for a classical particle. Of course, in the classical limit $(\hbar\to 0)$ the correlations for $X_Q$ and $P_Q$ coincide with the classical correlations for $X_{cl}$ and $P_{cl}$.

If we want to describe a quantum particle in our formalism based on the probability density in phase space we have to modify the fundamental evolution law for $w(x,p)$. We will do so by postulating in the evolution equation for the classical wave function \eqref{89A} a different Hamiltonian, namely 
\be\label{GP2}
H_W=-i\frac pm\partial_x+V\left(x+\frac i2\partial_p\right)-V
\left(x-\frac i2\partial_p\right),
\ee
such that the new fundamental evolution law for the probability density $w$ is specified by 
\be\label{J2}
i\partial_t\psi_C(x,p)=H_W\psi_C(x,p).
\ee
In consequence, the classical probability density $w(x,p)$ obeys a new non-linear time evolution equation instead of the Liouville equation. 
\ba\label{Q11}
\partial_t w(x,p)&=&-2\sqrt{w(x,p)}L_W\sqrt{w(x,p)},\nn\\
L_W&=&iH_W.
\ea
We note that $L_W$ is a real operator. The evolution equation \eqref{Q11} is of the general type \eqref{N1} discussed in the introduction. 

This new fundamental dynamical equation reduces to the Liouville equation in the classical limit $\hbar\to 0$, as can be seen easily by restoring the $\hbar$-factors in the fundamental evolution law
\ba\label{Q12}
\partial_t w&=&-2\sqrt{w}L_W\sqrt{w},\\
L_W&=&\frac pm\partial_x+iV\left(x+\frac{i\hbar}{2}\partial_p\right)-iV
\left(x-\frac{i\hbar}{2}\partial_p\right).\nn
\ea
We observe $L_W(\hbar\to 0)=\hat L$. It is straightforward to verify that with eq. \eqref{J2} the time evolution of $\rho_w$ obeys the usual time evolution for the density matrix of a quantum particle,
\be\label{Q13}
\partial_t\rho_w=
\left\{-\frac pm\partial_x-iV\left(x+\frac i2\partial_p\right)+iV
\left(x-\frac i2\partial_p\right)\right\}\rho_w.
\ee
We emphasize that in our approach the time evolution \eqref{Q13} of $\rho_w$ is a {\em consequence} of the fundamental evolution law \eqref{Q12}.

In summary, the expectation values of the quantum observables $X_Q$ and $P_Q$ and all their correlation functions obey all the relations for a quantum particle in a potential, including their time evolution. Starting at some initial time $t_0$ with a classical probability distribution which corresponds to a given $\rho_w(x,p)$, all quantum laws for a quantum  particle in a potential are obeyed for all times, including characteristic phenomena as interference and tunneling. These correlation functions are the only thing measurable in this system - demonstrating that quantum mechanics can be described in terms of a classical probability distribution in phase space.

\medskip\noindent
{\bf 3. \quad Classical and quantum probabilities}

The probability to find a particle at the quantum position $x$ is given by
\be\label{Q14}
w_Q(x)=\int_p\rho_w(x,p),
\ee
implying
\be\label{171A}
\kl f(X_Q)\kr=\int_xf(x)w_Q(x).
\ee
In terms of the classical wave function $\psi_C(x,p)$ one finds
\be\label{171B}
w_Q(x)=\int_{r,q,q'}\psi_C\left(x+\frac r2,q\right)\psi_C\left(x+\frac r2,q'\right)\cos 
\big[(q'-q)r\big].
\ee
This differs from the probability to find the particle at the classical position $x$, given by
\be\label{Q15}
w_C(x)=\int_p\psi^2_C(x,p)=\int_pw(x,p),
\ee
with
\be\label{171C}
\kl f(X_{cl})\kr=\int_xf(x)w_C(x).
\ee
We may write $w_Q(x)$ in the form
\ba\label{171D}
w_Q(x)=\int_{z,z',q,q'}\psi_C(z,q)\tilde L_Q(x;z,z',q,q')
\psi_C(z',q')
\ea
with $\tilde L_Q(x)$ the operator indicating if the particle is present at position $x$ or not. ($\tilde L_Q$ stands here for a ``location observable'' and should not be confounded with the Liouville operator or other operators for the time evolution.) For the classical position one has
\be\label{171E}
\tilde L_C(x;z,z',q,q')=\delta(x-z)\delta(z-z')\delta(q-q'),
\ee
while for the quantum quantum position one finds
\be\label{171F}
\tilde L_Q(x;z,z',q,q')=2\cos\big[2(q-q')(x-z)\big]\delta(z-z').
\ee
We observe that $\tilde L_Q$ also involves locations $(x-z)$ different from $x$. Furthermore, $\tilde L_Q$ is no longer a diagonal operator in momentum space. 

\medskip\noindent
{\bf 4. \quad Classical and quantum observables}

On the level of the classical wave function we can define both quantum  and classical observables for position and momentum. They obey the commutation relations 
\ba\label{173A}
[X_{cl},X_Q]&=&0~,~[P_{cl}, P_Q]=0,\nn\\
~[X_{cl},P_Q]&=&\frac{i}{2}~,~[X_Q,P_{cl}]=\frac i2.
\ea
We can express both the classical Hamiltonian $H_L$ and the quantum Hamiltonian $H_W$ in terms of the associated operators
\ba\label{173B}
H_L=\frac 2m P_{cl}(P_Q-P_{cl})+2V'(X_{cl})(X_Q-X_{cl}),\nn\\
H_W=\frac 2m P_{cl}(P_Q-P_{cl})+V(X_Q)-V(2X_{cl}-X_Q).
\ea
This yields the general commutator relations 
\ba\label{P3}
i[H_L,X_{cl}]&=&\frac{1}{m}P_{cl}~,~i[H_L,P_{cl}]=-V'(X_{cl}),\nn\\
i[H_W,X_Q]&=&\frac 1m P_Q~,~i[H_W,P_Q]=-V'(X_Q).
\ea
Using the Heisenberg picture the time evolution of the classical observables $X_{cl},P_{cl}$ with Hamiltonian $H_L$ is the same as the time evolution of the quantum observables $X_Q,P_Q$ with Hamiltonian $H_W$. For the special case of a harmonic potential the classical and quantum Hamiltonians coincide, $H_L=H_W$, and the time evolution of the probability density in phase space is identical.

\section{Correlations and incomplete statistics}
\label{Correlations and incomplete statistics}

Having realized quantitatively all dynamical features of a quantum particle within a classical statistical ensemble one may address some of the conceptual questions of quantum mechanics. Several of them are actually very similar to questions that arise within a statistical treatment of classical collective systems as effective one-particle states. The central issue concerns the coarse graining of the information that occurs in the transition from a description in terms of microstates or substates (say an ensemble for the molecules in the water droplet) to a subsystem for the effective one-particle description (say the probability density $w(x,p)$ for the water droplet). The probability density for the positions and momenta of $10^{20}$ water molecules contains much more information than the function $w(x,p)$ for the collective droplet. In the process of coarse graining of the information one therefore discards a great amount of the conceptually possible information on the substate level. One has to think which statistical quantities ``survive'' this loss of information. In particular, classical correlation functions, which are defined for the ensemble of microstates, may not carry over to the ensemble for the subsystem, leading to ``incomplete statistics'' \cite{3}, \cite{CWAA1}. This notion of incomplete statistics is crucial for an understanding of the correlations in quantum systems.

Indeed, the water droplet can serve as an illustration for several of the issues encountered by an embedding of quantum mechanics into classical statistics. The precise definition of the ``substates'' is not important in this context. They may be related to internal degrees of freedom as deformation modes on a hydrodynamic level, or properties of individual molecules on the level of a microscopic description of liquids. 

The first conceptual point concerns the role of observables on the level of an effective one-particle description and the issue of incomplete statistics. ``One-particle observables'' are those for which expectation values can be computed unambiguously from the classical wave function $\psi_C(x,p)$. Two different classical observables $A_1$ and $A_2$ on the substate level may have identical expectation values for all possible one-particle wave functions $\psi_C(x,p)$ and associated probability distributions $w(x,p)$. From the point of view of the effective ``one-particle subsystem'' they are identical observables. All classical observables $A_n$ with this property may be grouped into an associated equivalence class of one-particle observables $A$. Measurements in the subsystem cannot resolve between different members of the equivalence class. Nevertheless, the observables $A_1$ and $A_2$ can be different classical observables for a more microscopic description of the droplet. If the information concerning the internal degrees of freedom of the droplet can be resolved, the observables $A_1$ and $A_2$ can yield different values for these internal degrees of freedom. From the point of view of the one-particle subsystem the unresolved internal degrees of freedom are part of the ``environment. The notion of ``environment'' is taken here in the sense of information not employed and not available for the description of the subsystem. As evident for the water droplet, this does not involve necessarily a separation between subsystem and environment in space, or in the sense of no causal influence. 

The notion of equivalence classes of classical observables for the description of properties of the one-particle subsystem has important consequences for the choice of correlation functions that describe a sequence of two measurements of properties of the subsystem. Let us consider a further observable $B$ (not in the equivalence class $A$). The microscopic classical correlation functions $A_1\cdot B$ and $A_2\cdot B$ are often different. This difference concerns properties of the environment. In other words, if the substate observables $A_1, A_2$ and $B$ can all be mapped to one-particle observables whose expectation value can be computed from $\psi_C(x,p)$, this does not imply that the substate observables $A_1\cdot B$ and $A_2\cdot B$ can also be mapped to one particle observables. Explicit simple examples where this is not possible can be found in ref. \cite{CWAA1}.

Under such circumstances the classical correlations $A_1\cdot B$ and $A_2\cdot B$ cannot be computed anymore from $\psi_C(z,p)$ - the absence of this information is the meaning of incomplete statistics. In contrast, an idealized sequence  of two measurements of subsystem properties should only depend on the equivalence classes for $A$ and $B$. It should therefore be described by a correlation function that differs from the classical correlation function. One can show \cite{CWAA1} that quantum observables are indeed associated to such equivalence classes, and that quantum correlations are compatible with the structure of equivalence classes. The association of a quantum observable with a whole equivalence class of classical observables avoids conflicts with the Kochen-Specker theorem \cite{KS}, as demonstrated explicitly in ref. \cite{CWAA1}.

The property of incomplete statistics \cite{3}, \cite{CWAA1} where the measurement correlation is not based on joint probabilities, is a key point \cite{CW2}, \cite{CWAA1} for an understanding of quantum mechanics. Indeed, complete statistical systems, for which the measurement correlation employs the joint probabilities, have to obey Bell's inequalities \cite{BS}. The experimental verification of a violation of Bell's inequalities is in agreement with our argument that joint probabilities cannot be used for a sequence of ideal measurements in the one-particle subsystem. One may conceive the embedding of quantum mechanics into classical statistics within a fundamental probabilistic setting \cite{GenStat} of probabilistic realism. On the level of microscopic states joint probabilities may be available. However, on the level of the one-particle subsystem a description in terms of complete statistics is typically no longer possible \cite{CWAA1}. For incomplete statistics the EPR-paradoxon \cite{EPR} can be resolved satisfactorily \cite{CWAA1}, \cite{CW2}. 

In the setting of sects. \ref{Classical wave function}, \ref{Quantumparticlesfrom} the one-particle state is characterized by the ``classical wave function'' $\psi_C(x,p)$. This contains the information that specifies the statistical subsystem. (We will see in sect. \ref{Quantum density matrix} that this information can be reduced further in case of a quantum particle.) We emphasize the genuinely probabilistic concept of a one-particle state. In general, it is characterized by a distribution of values of the observables $X_{cl}$ and $P_{cl}$, rather than by the sharp values for a ``classical trajectory state''. One-particle states for which the statistical observables of the type $X^2_s$ or $P^2_s$ are well defined cannot have an arbitrary sharp distribution. For example a ``classical trajectory state'', where $w(x,p)$ has support only for one particular point $(x,p)$ in phase space, does not admit the statistical observables $X^2_s$ and $P^2_s$. The notion of a one-particle state, characterized by $\psi_C(x,p)$, should not be confounded with the states of the classical statistical ensemble, characterized by points in phase space $(x,p)$. 

For the observables $X_{cl}$ and $P_{cl}$ for classical position and momentum the joint probabilities are still available on the level of the one-particle subsystem. Indeed, the joint probability to find the value $x$ for the observable $X_{cl}$, and $p$ for the observable $P_{cl}$, is precisely given by $w(x,p)$. ``Classical correlations functions'' as $\kl X_{cl}\cdot P_{cl}\kr$ are well defined for these observables. In contrast, the observables $X_Q$ and $P_Q$ for quantum position and momentum do not admit joint probabilities. One may want to represent the ``quantum observables'' $X_Q$ and $P_Q$ as classical statistical observables on a substate level \cite{CWAA}. On this substate level the classical products $X_Q\cdot P_Q$ can be well defined classical observables for some specific choice of substate observables $X_Q$ and $P_Q$. (Recall that this choice is not unique.) However, the classical product $X_Q\cdot P_Q$ is not among the one-particle observables. It is not computable in terms of the wave function $\psi_C(x,p)$. In particular, the observables $X_Q$ and $P_Q$ have no values that can be associated to a point in phase space $(x,p)$. A classical product $X_Q\cdot P_Q$ with fixed values for a phase space point $(x,p)$ cannot be defined either. One can find probability distributions or classical wave functions for which either $X_Q$ or $P_Q$ has a fixed value, but not both simultaneously. Also on this level there is no meaning of a classical product $X_Q\cdot P_Q$. On the other hand, the quantum products $X_QP_Q$ and $P_QX_Q$ are well defined in the usual sense of multiplication of operators. They are computable from the classical wave function $\psi_C(x,p)$ and can be used for measurement correlations describing a sequence of two measurements in the one-particle subsystem. 

A further important property of quantum observables is visible in the one-particle picture of a water droplet. On the microscopic level certain observables may only take discrete values. According to the rules of classical statistics these values in the spectrum of the classical observable are the only possible outcomes of measurements. An observable remains a one-particle observable if its expectation value can be computed from the information contained in the classical wave function $\psi_C(x,p)$. However, it does not necessarily take a fixed value in every one-particle state. In general, on the level of the one-particle subsystem the observables become probabilistic observables \cite{CW2}, \cite{PO}.  For any state $\psi_C(x,p)$ of the subsystem they have a probabilistic distribution of the values in their spectrum. This issue becomes particularly important if we also want to describe discrete internal properties of a particle, for example its spin in a given direction \cite{CW2,CWE,CW1}.

\section{Quantum density matrix}
\label{Quantum density matrix}

Since we have described the probability distribution in phase space and its time evolution is terms of a wave function, we can use all the familiar techniques of the quantum formalism. In particular, we can change the basis of the Hilbert space. We will use this in order to extract from the classical wave function $\psi_C(x,p)$ the quantum density matrix $\rho_Q(x,x')$. This further coarse graining of the information maps the probability distribution for the classical statistical ensemble for a one-particle state to the familiar formalism of quantum mechanics. 

\medskip\noindent
{\bf 1. \quad Position basis}

A useful basis is the ``position basis'' where the wave function,
\be\label{209}
\tilde \psi_C(x,y)=\int_p e^{ip(x-y)}\psi_C\left(\frac{x+y}{2},p\right)=\tilde\psi^*_C(y,x),
\ee
obtains by a Fourier transform from the ``phase space basis'' $\psi_C(z,p)$. (In general, $\tilde\psi_C(x,y)$ is a complex function.) In this basis the quantum Hamiltonian takes a particularly simple form 
\ba\label{210}
H_W=H_Q-\tilde H_Q~,~[H_Q,\tilde H_Q]=0,
\ea
with
\ba\label{211}
H_Q&=&-\frac{1}{2m}\partial^2_x+V(x),\nn\\
\tilde H_Q&=&-\frac{1}{2m}\partial^2_y+V(y).
\ea

The classical wave function associated to a pure state quantum mechanical wave function $\psi_Q(x)$ reads in this basis
\be\label{212}
\tilde\psi_C(x,y)=\psi_Q(x)\psi^*_Q(y).
\ee
As is well known, the quantum mechanical Schr\"odinger equation for $\psi_Q(x)$ with Hamiltonian $H_Q$ \eqref{211} translates to the von-Neumann equation for the quantum density matrix $\rho_Q(x,x')$, and by a partial Fourier transform to eq. \eqref{Q13} for the Wigner function. With eq. \eqref{212} the wave function $\tilde\psi_C(x,y)$ coincides with the quantum density matrix $\rho_Q(x,y)$. In turn, we find for pure quantum states the simple relation
\be\label{Q19}
\rho_w(x,p)=\psi_C(x,p).
\ee
One may verify that for the special case of a quantum pure state the general definition of $\rho_w$ by the folding \eqref{AA1} agrees with eq. \eqref{Q19}. Eq. \eqref{Q19} does not hold for mixed quantum states. 

It is a characteristic property of the quantum Hamiltonian $H_W$ \eqref{173B} that it can be decomposed into two commuting pieces \eqref{210}, with
\ba\label{215A}
H_Q=\frac{1}{2m}P^2_Q+V(X_Q)~,~\tilde H_Q=\frac{1}{2m}\tilde P_Q^2+V(\tilde X_Q),
\ea
where
\ba\label{215B}
\tilde X_Q=2X_{cl}-X_Q~,~\tilde P_Q=P_Q-2P_{cl},
\ea
with commutators
\ba\label{213}
~[\tilde X_Q,\tilde P_Q]&=&i,\\
~[\tilde X_Q,X_Q]&=&[\tilde X_Q,P_Q]=[\tilde P_Q,X_Q]=
~[\tilde P_Q,P_Q]=0.\nn
\ea
In terms of the statistical observables $X_s,P_s$ one has
\be\label{214}
\tilde X_Q=X_{cl}-\frac12 X_s~,~\tilde P_Q=-\left(P_{cl}-\frac12 P_s\right).
\ee

We observe that a reflection of the phase space momentum $p\to -p$ results in
\be\label{215}
X_s\to-X_s~,~P_{cl}\to-P_{cl},
\ee
while $X_{cl}$ and $P_s$ remain invariant. This reflection therefore induces
\be\label{216}
X_Q\to \tilde X_Q~,~P_Q\to\tilde P_Q~,~H_Q\to \tilde H_Q~,~H_W\to -H_W.
\ee
The quantum time evolution is invariant under the time reflection $t\to-t~,~p\to -p$. Eq. \eqref{210} constitutes a ``decomposition property'' of the quantum Hamiltonian, which can be written as the difference of two commuting pieces which are mapped into each other by time reflection with associated momentum reflection. One may postulate the decomposition property as a fundamental characteristics of the time evolution of the probability distribution which describes a quantum particle. In general, this property is not realized for the classical time evolution which uses $H_L$.

\medskip\noindent
{\bf 2. \quad Coarse graining}

From the classical wave function \eqref{209} we can construct the ``classical density matrix'' in the usual way,
\be\label{217}
\tilde\rho_C(x,x',y,y')=\tilde\psi_C(x,y)\tilde\psi^*_C(x',y')=\tilde\psi_C(x,y)
\tilde\psi_C(y',x').
\ee
We may proceed to a ``coarse graining'' by ``integrating out the $y$-variable'' or ``taking a subtrace''
\be\label{218}
\rho_Q(x,x')=\int_y\tilde\rho_C(x,x'y,y).
\ee
As it should be, the coarse grained density matrix $\rho_Q$ obeys the properties of a quantum mechanical density matrix
\be\label{219}
\rho_Q^*(x',x)=\rho_Q(x,x')~,~\int_x\rho_Q(x,x)=1.
\ee
The quantity $\tilde\rho_C(x,x',y,y')$ is by construction a positive matrix. The coarse graining preserves this positivity such that $\rho_Q(x,x')$ is a positive matrix with all eigenvalues positive or zero. 

The quantum position and momentum observables $X_Q,P_Q$ do not involve the variable $y$, such that expectation values can be computed from $\rho_Q$ by the usual quantum rule
\ba\label{220}
&&\kl F(X_Q,P_Q)\kr =\tr \{F(X_Q,P_Q)\rho_Q\},\nn\\
&&X_Q=x~,~P_Q=-i\partial_x.
\ea
The decomposition property of the quantum time evolution implies the von-Neumann equation for the time evolution of $\rho_Q$
\be\label{221}
\partial_t\rho_Q=-i[H_Q,\rho_Q].
\ee
This holds for an arbitrary classical wave function $\psi_C(x,p)$ or phase-space probability $w(x,p)$, both for situations describing pure or mixed quantum states. Finally, the Wigner transform of $\rho_Q$ is given by $\rho_w$ in eq. \eqref{AA1}. This is the origin of the definition by this particular folding. We have discussed the association of a quantum particle with a coarse grained classical probability distribution in ref. \cite{3A}.

\section{Quantum and classical energy}
\label{Quantum and classical energy}

The notion of energy is very different for quantum and classical particles. This holds even for a harmonic potential. We define the classical and quantum energy by
\be\label{P5}
H_Q=\frac{1}{2m}P^2_Q+V(X_Q)~,~H_{cl}=\frac{1}{2m}P^2_{cl}+V(X_{cl}).
\ee
One verifies that the quantum energy is conserved by the quantum evolution
\be\label{Q16}
[H_W,H_Q]=0,
\ee
while the classical energy is conserved by the classical evolution
\be\label{Q17}
[H_L,H_{cl}]=0.
\ee
For the special case of a harmonic oscillator, where $H_W=H_L$, both $H_Q$ and $H_{cl}$ are conserved. 

The two energy observables do not commute, however,
\be\label{P6}
[H_Q,H_{cl}]=\frac{1}{2m}
[P^2_Q,V(X_{cl})]-\frac{1}{2m}[P^2_{cl},V(X_Q)],
\ee
where we may use the general relations
\ba\label{P7}
[P^2_Q,V(X_{cl})]&=&-\frac14 V''(X_{cl})-iV'(X_{cl})P_Q,\nn\\
~[P^2_{cl},V(X_Q)]&=&-\frac14 V''(X_Q)-iV'(X_Q)P_{cl}.
\ea

\medskip\noindent
{\bf 1.\quad Harmonic potentials}

Let us consider first the special case of a harmonic potential in one space dimension
\be\label{Q18}
V(x)=\frac c2 x^2.
\ee
For the harmonic oscillator the eigenstates of $H_Q$ have the usual equidistant discrete spectrum, which follows from the operator algebra. Indeed, we can introduce the quantum mechanical eigenstates $\psi_{Q,n}(x)$ in a basis where $X_Q\psi_Q=x\psi_Q$. In this basis the eigenfunctions of $H_Q$ obey
\ba\label{P8}
H_Q\psi_{Q,n}(x)&=&E_n\psi_{Q,n}(x)=\omega\left(n+\frac 12\right)\psi_{Q,n}(x),\nn\\
\omega&=&\sqrt{\frac cm}.
\ea
The classical wave function which corresponds to the quantum mechanical eigenstate of $H_Q$ is given by
\be\label{P9}
\psi_{C,n}(z,p)=\int_r e^{-ipr} \psi_{Q,n}\left(z+\frac r2\right)\psi^*_{Q,n}\left(z-\frac r2\right),
\ee
where we use now $z=(x+y)/2$ as argument of the classical probability distribution and wave function. In particular, the ground state of the quantum  harmonic oscillator
\be\label{P10}
\psi_{Q,0}(x)=\left(\frac{m\omega}{\pi}\right)^{1/4}\exp \left(-\frac{m\omega x^2}{2}\right)
\ee
corresponds to a static classical wave function $\psi_{C,0}(z,p)$ and probability distribution $w_0(z,p)$ given by
\ba\label{P11}
\psi_{C,0}(z,p)&=&2\exp (-m\omega z^2)\exp \left(-\frac{p^2}{m\omega}\right),\nn\\
w_0(z,p)&=& 4\exp(-2m\omega z^2)\exp\left(-\frac{2p^2}{m\omega}\right)\nn\\
&=&4\exp\left\{-\frac 4\omega\left(\frac{p^2}{2m}+V(z)\right)\right\}.
\ea

In contrast, the classical energy has a continuous spectrum of positive energies, $\epsilon\geq V_{\rm min}=0$,
\be\label{P12}
H_{cl}\psi(z,p;\epsilon)=\epsilon\psi(z,p;\epsilon).
\ee
We may concentrate on static probability distributions with $\kl P_{cl}\kr =0$, as 
\be\label{P13}
w_\epsilon(z,p;\epsilon)=\omega\delta\left(\frac{p^2}{2m}+V(z)-\epsilon\right).
\ee
According to the Liouville equation this probability distribution is static. The expectation values of the classical kinetic and potential energy obey the virial theorem
\be\label{P14}
\frac{1}{2m}\kl P^2_{cl}\kr=\kl V(X_{cl})\kr=\frac{\epsilon}{2}.
\ee
The energy dispersion vanishes for eq. \eqref{P13}
\be\label{P15}
\kl H^2_{cl}\kr-\kl H_{cl}\kr^2=\int_{z,p}
\left(\frac{p^2}{2m}+V(z)\right)^2w_\epsilon(z,p)-\epsilon^2=0.
\ee

We emphasize that the vanishing classical energy dispersion is not sufficient to fix the classical probability distribution uniquely (in contrast to the quantum energy). Another example with a vanishing energy dispersion is given by
\be\label{P16}
w(z,p;\epsilon)=\frac{\pi}{L}
\theta \big(\epsilon-V(z)\big)\Big(\delta\big (p-\hat p(z)\big )
+\delta\big (p+\hat p(z)\big )\Big)
\ee
where we use
\ba\label{P17}
\hat P(z)&=&\sqrt{2m\big (\epsilon-V(z)\big)}~,~V(z)=\frac c2 z^2,\nn\\
L&=&\sqrt{8\epsilon/c}.
\ea
For this probability distribution the expectation values of classical kinetic and potential energy do not obey the virial theorem
\be\label{P18}
\frac{1}{2m}\kl P^2_{cl}\kr=2 V(X_{cl})\kr=\frac{2\epsilon}{3}.
\ee
The distribution \eqref{P16} is not static.

\medskip\noindent
{\bf 2.\quad Static probability distributions}

Static probability distributions according to the Liouville equation can be obtained if $w(z,p)$ depends on the phase-space coordinates only through the combination $E(z,p)=p^2/2m+V(z),w(z,p)=w\big(E(z,p)\big)$, or if the associated classical wave functions take this form
\be\label{P20}
\psi_C(z,p)=\psi\big (E(z,p)\big).
\ee
They can be written as superpositions of the $w_\epsilon$ given by eq. \eqref{P13}, or the associated wave functions $\psi_\epsilon=\sqrt{w_\epsilon}$,
\be\label{P21}
\psi_C(z,p)=\int^\infty_0 d\epsilon F(\epsilon)\psi_\epsilon(z,p).
\ee
The quantum eigenstates $\psi_{C,n}(z,p)$ are static states of this form. It is an interesting question if these states are somehow singled out, in particular the quantum ground state \eqref{P11}.

The most general static classical wave function is an eigenstate of $H_L$ with eigenvalue $0$,
\be\label{P22}
H_L\psi_C(z,p)=0.
\ee
It is necessarily of the form \eqref{P20}, $\psi_C=\psi(E)$, normalized according to 
\be\label{P23}
\int^\infty_0 dE\psi^2(E)=\omega.
\ee
Superpositions of static classical wave functions remain static. We can expand the most general static classical wave function in terms of eigenstates of the classical energy according to eq. \eqref{P21}, with 
\be\label{P24}
\psi_\epsilon(E)=\sqrt{\omega}\delta^{1/2}(E-\epsilon),
\ee
where $\delta^{1/2}$ may be defined by an appropriate limit of a Gauss function. Alternatively, one may choose an expansion in terms of the eigenfunctions $\psi_n(E)$ of the quantum energy
\be\label{P25}
\psi(E)=\sum^\infty_{n=0}c_n\psi_n(E).
\ee
At this stage nothing seems to single out the discrete quantum basis $\psi_n(E)$ as compared to the continuous classical basis $\psi_\epsilon (E)$. 

On the level of the probability distributions in phase space or the classical wave functions we have a continuity of static distributions, namely all $\psi(E)$ normalized according to eq. \eqref{P23}. The eigenstates of the quantum energy, $\psi_n(E)$ form a discrete subset. The stability properties of small fluctuations around a given static state $\bar\psi(E)$ do not single out the ``quantum states'' $\psi_n(E)$ either. Since the time evolution equation is linear and $H_L\bar\psi(E)=0$, the time evolution of a fluctuation $\delta\psi(z,p)=\psi(z,p)-\bar\psi(E)$ remains independent of the choice of $\bar\psi(E)$. What singles out the quantum states $\psi_n(E)$ is the additional property that they are eigenstates of the quantum energy
\ba\label{200}
H_Q\psi_{C,n}(z,p)&=&\left\{\frac{1}{2m}\left(p-\frac i2\partial_z\right)^2\right.\nn\\
&&\left. +V\left(z+\frac{i}{2}\partial_p\right)\right\}\psi_{C,n}(z,p)\nn\\
&=&\left(n+\frac{1}{2}\right)\omega\psi_{C,n}(z,p).
\ea
We may conjecture that small deviations from a harmonic potential or the interaction with the environment are responsible for the special role of the quantum states observed in Nature.

\medskip\noindent
{\bf 3.\quad Unharmonic potentials}

For unharmonic potential the classical and quantum time evolution of the probability density no longer coincide, $H_L\neq H_W$. For the classical time evolution the classical energy is conserved, while for the quantum time evolution the quantum energy is conserved,
\be\label{201}
[H_L,H_{cl}]=0~,~[H_W,H_Q]=0.
\ee
If the fundamental evolution law for the probability distribution in phase space is given by the quantum evolution \eqref{J2} it seems therefore natural that the eigenstates of $H_Q$ are singled out by the dynamics.

For the classical time evolution the most general static wave function is again given by $\psi(E)$, normalized now according to 
\ba\label{202}
&&\int^\infty_{E_{min}} dE f_N(E)\psi^2(E)=1,\nn\\
&&f_N(E)=\sqrt{\frac{m}{2\pi^2}}\int_x\big (E-V(x)\big)^{-1/2},
\ea
where the $x$-integral is bounded to $V(x)\leq E$. (For the harmonic oscillator $f_N(E)=\omega^{-1}$.) 

The static states for the quantum time evolution are more complicated. First we may show that all pure quantum states which correspond to eigenvalues to $H_Q~,~H_Q\psi_{Q,n}(x)=E_n\psi_{Q,n}(x)$, lead to static classical wave functions according to the construction \eqref{P9},
\be\label{203}
H_W\psi_{C,n}(z,p)=0.
\ee
Linear combinations
\be\label{204}
\bar\psi_C(z,p)=\sum_n q_n\psi_{C,n}(z,p)
\ee
are also static. With
\be\label{205}
H_Q\psi_{C,n}(z,p)=E_n\psi_{C,n}(z,p)
\ee
we find the average quantum energy in such a static state
\be\label{206}
\kl H_Q\kr=\sum_n q^2_n E_n,
\ee
provided the states $\psi_{C,n}$ are orthogonal,
\be\label{207}
\int_{z,p}\psi_{C,n}(z,p)\psi_{C,n}(z,p)=\delta_{nm}.
\ee
In this case we may associate $q^2_n$ with a probability $p_n$ to find the state $\psi_{C,n}$, since the normalization of $\bar\psi(z,p)$ in eq. \eqref{204} requires
\be\label{208}
\sum_n p_n=\sum_n q^2_n=1.
\ee
In particular, the minimal value for $\kl H_Q\kr$ is obtained for the quantum mechanical ground state, $p_0=1,p_{n\neq 0}=0$. 

\medskip\noindent
{\bf 4. \quad Stability of atoms}

In summary, the quantum time evolution of the probability distribution in phase space is characterized by a conserved quantum energy $H_Q$. This is bounded from below if the potential has the appropriate properties (the same as for quantum mechanics). The minimal value of $H_Q$ corresponds to the quantum mechanical ground state. The associated probability distribution in phase space is static. If the particle can exchange quantum energy with its environment - as the electron in an atom by radiation - it is plausible that often the asymptotic state of the time evolution (including the exchange with the environment) is given by the state with the minimal quantum energy. This clearly singles out the quantum ground state among all possible static probability distributions. In other words, we have found a possible explanation for the identity of atoms in a purely classical statistical framework! Of course, The modification of the time evolution of the classical wave function by replacing $H_L\to H_W$ is crucial for this purpose. It would be very interesting to find out if similar properties carry over to zwitters if collective states are described by an evolution law close to a quantum particle.

\section{Continuous interpolation between quantum and classical particle}
\label{Continuousinterpolation}

We have described a classical particle and a quantum particle within the same formalism. Both concepts find a formulation within a classical statistical ensemble. A convenient way to display the information contained in the classical probability density $w(x,p)$ is the classical wave function $\psi_C(x,p)$, with $w=\psi^2_C$. The use of the classical wave function permits us to employ the same quantum formalism for classical and quantum particles. It is also convenient to highlight the main differences between classical and quantum particles: the use of different observables for position and momentum and a different time evolution of the probability density in case of unharmonic potentials. Both differences vanish in the classical limit that may formally be taken as $\hbar\to 0$. 

It is possible to interpolate continuously between the quantum particle and the classical particle. Observables that interpolate between the quantum observables and classical observables correspond to the possibility of ``sharpened measurements''. On the other hand, we will denote particle like objects that realize an interpolated time evolution as ``zwitters'' - they are neither purely quantum nor purely classical \cite{3A}. It will be interesting to find out which objects and measurements can be well approximated by zwitters or sharpened measurements. The mathematical construction is free of contradictions, but the coarse graining of a real collective system to an effective one-particle subsystem may be rather complicated in practice. Thus the question of the existence of zwitters may partly become an experimental issue. In particular, experiments can test quantitatively how pure is the quantum behavior of particles in a given experimental setting. This holds not only for effective one particle states but also on a fundamental level. 

Interpolations are possible both for the question which observables are appropriate and for the time evolution of the probability density. For a free particle or a particle in a harmonic potential the classical and quantum Hamiltonians coincide. The distinction between classical and quantum behavior reduces then to the choice of the appropriate observables, with the possibility of a continuous interpolation.

The interpolation between a quantum and a classical particle is not unique. There are many different possibilities to choose interpolating observables or interpolating Hamiltonians. It seems likely to us that the choice of a specific interpolation is not an issue of fundamental physics. As we have discussed in the introduction, any collective state that admits an effective one-particle description with an evolution equation \eqref{N1} that differs from the two extreme cases of a classical particle or a quantum particle can be considered as a zwitter. A one-particle picture of a water droplet will not behave as a classical particle.

The difference between a classical point particle and an effective one-particle description of a droplet results from the unresolved degrees of freedom in the second case. For example, we may choose the center of mass coordinate and the total momentum  of the droplet in order to characterize the one-particle states $(x,p)$. With $m$ the mass of the droplet we can define the kinetic energy $E_{kin}=p^2/(2m)$. For a droplet in a static external potential we may also define a potential energy by evaluating the potential at the position of the center of mass $V(x)$. However, $H_{cl}=E_{kin}+V(x)$ is not conserved, since it is possible to transfer energy from the internal degrees of freedom to $H_{cl}$ and vice versa, for example by a deformation of the droplet. If the potential has a local maximum with height $V_0$ and the initial energy $H_{cl}$ is smaller than $V_0$, a classical point particle cannot cross such a barrier. In contrast, this is not excluded for a droplet since $H_{cl}$ is not conserved. The passing of a barrier under suitable circumstances resembles the tunneling in quantum mechanics. Since the Liouville equation implies a conserved $H_{cl}$ we infer that for a droplet the time evolution of the probability distribution $w(x,p)$ cannot obey the Liouville equation. The droplet does not behave as a classical point particle but rather resembles a zwitter. Further differences to a classical point particle arise if the potential is not static but only stationary in a statistical sense that the ensemble average does not depend on time. 

While it is rather obvious that the droplet-dynamics deviates from a classical point particle it is not guaranteed that it behaves as a zwitter. A zwitter requires an effective one-particle description in the sense of eq. \eqref{N1}. (For example, it is possible that the description of a droplet needs additional stochastic terms in the evolution equation, or a realistic approximation may involve additional degrees of freedom beyond $(x,p)$.) Within the setting \eqref{N1} we have specialized to evolution equations that are linear in the classical wave function $\psi_C$. Furthermore, in the setting of this paper the influence of the environment is encoded in a function $V(x)$, and the specification how the dynamics is influenced by the presence of $V(x)$. While all this is not the most general setting, we find it rather likely that this specific restricted form of zwitters can actually be realized in suitable experimental settings or physical situations, at least to a good approximation. In contrast to our description of a microscopic quantum particle there is no need, however, to identify the parameter $\hbar$ in eq. \eqref{Q12} - or zwitter-generalizations of it - with Planck's constant. For a description of collective states by zwitters $\hbar$ will play the role of an effective macroscopic or mesoscopic constant, which is adapted to the particular situation.

On the other hand, for microscopic particles it is well conceivable that there are profound reasons for the quantum evolution law \eqref{Q12} and the choice of quantum observables \eqref{Q9} in terms of a more fundamental quantum field theory. In ref. \cite{CWPT} we have obtained the quantum  formalism for a multi-particle theory from a probabilistic formulation of time within the framework of a classical statistical ensemble. In refs. \cite{CWF,CDW} we have formulated time evolution laws for classical statistical ensembles that describe the dynamics of quantum field theories for fermions. In such a setting the time evolution of the probability distribution for a quantum particle discussed in this paper describes the idealized situation of a perfectly isolated single particle. Such a scenario explains why $\hbar$ is universal. In particular, we have proposed in ref. \cite{CDW} an evolution equation for a classical statistical ensemble of Ising spins that is equivalent to a quantum field theory for Dirac fermions in an arbitrary electromagnetic field. The non-relativistic approximation for one-particle states yields the Schr\"odinger equation for a quantum particle in a potential. In this model $\hbar$ appears as a pure conversion constant of units and can be identified with Planck's constant. In many parts of the present paper we set $\hbar=1$ such that distance and inverse momentum units can be converted into each other using the appropriate microscopic or effective macroscopic value for $\hbar$. 

Within a fundamental quantum field theory zwitters can describe deviations from a perfectly isolated one-particle quantum state. One may associate the quantum subsystem with the information contained in the quantum density matrix. A zwitter-Hamiltonian  could then describe the lack of complete isolation. Indeed, a zwitter-Hamiltonian will result in a lack of unitarity of the time evolution of $\rho_Q$. In this case a zwitter-Hamiltonian can describe the phenomena of decoherence \cite{DC} or syncoherence \cite{CW2}. We will find in sect. \ref{Zwitters - continuous} that the ground state energy of an atom acquires a nonzero width in case of a zwitter-Hamiltonian. Again, this could be interpreted as a lack of isolation.

Many effective time evolution equations for a quantum density matrix in case of imperfect isolation have been investigated in the past, for example by coupling the quantum particle to a heat bath and integrating out the degrees of freedom of the heat bath subsequently. The particularity of the zwitter particles is related to the fact that the quantum density matrix describes a subsystem of a particular ``extended subsystem''. This extended subsystem is given by the generalized particle, as described by the classical wave function $\psi_C(x,p)$. In turn, the generalized particle is treated as a perfectly isolated object. In particular, the time evolution of $\psi_C$ remains unitary for all choices of the zwitter-Hamiltonian. The lack of unitary appears only on the level of $\rho_Q$. 

Furthermore, the generalized particle is described by a pure state of the extended quantum formalism in phase-space. (Of course, one may generalize this setting by coupling the generalized particle to an environment or admitting mixed density matrices $\rho_C$ in phase space.) The embedding of the quantum subsystem into an extended isolated system, that can be described by phase-space probabilities, entails particularities for the lack of isolation of the quantum system. This distinguishes zwitters from a more general setting for incomplete isolation. It also justifies the use of the concept of zwitter particles, in contrast to particles coupled to an outside environment.

It would be interesting to find experimental settings that realize a zwitter particle in the context of a fundamental quantum theory. One may speculate that single atoms are characterized by an extremely small or vanishing interpolating parameter $\gamma$, such that the usual quantum description applies (almost) perfectly. Larger $\gamma$ could perhaps be expected for macro-molecules containing a very large number of atoms, or for condensates of ultracold atoms that can be treated as single quantum degree of freedom.

We finally emphasize the logical possibility that the interpolation is not universal, but that a universal lower bound for $\gamma$ could exist. In fact, in lowest order in $\gamma$ many possible interpolations coincide. It thus remains an experimental challenge to establish quantitative bounds for $\gamma$ even for the simplest quantum objects as the hydrogen atom. 

Summarizing this section, experiments testing zwitters can cover a wide range of physical phenomena, from classical collective systems to quantum collective systems, and from effective macroscopic one-particle states to tests of microscopic fundamental quantum laws. Among the many possibilities we will discuss in the following sections only a few particularly simple interpolations between classical and quantum. While some characteristic features of zwitters can be uncovered in this way, much richer interesting structures could perhaps be revealed in a more general treatment.

\section{Sharpened observables}
\label{Sharpened observables}
In this section we investigate the possibility that certain types of position or momentum measurements could be described by ``sharpened observables'' that differ from the usual quantum observables $X_Q$ and $P_Q$. One may focus the discussion of this question to quantum particles, i.e. assuming that the time evolution of the wave function in phase-space $\psi_C(x,p)$ is given by the quantum equation \eqref{GP2}, \eqref{J2}. A generalization to classical particles or zwitters is possible. The expression ``sharpened observables'' considers the interpolation between quantum and classical observables from the point of view of a quantum particle. From the point of view of a classical particle the same interpolation would correspond to ``smoothened observables''. 

The sharpened observables will involve the classical position and momentum observables $X_{cl}$ and $P_{cl}$. For quantum particles the expectation values of operators involving $X_{cl}$ and $P_{cl}$ can indeed be computed for a quantum pure state, due to the relation \eqref{212} which fixes the classical wave function in phase-space for any pure state quantum wave function $\psi_Q$. For mixed quantum states the relation between $\rho_Q$ and $\psi_C$ is not unique - different $\psi_C$ can lead to the same $\rho_Q$. In this case the measurement of sharpened observables could provide additional information about $\psi_C$ which is not contained in $\rho_Q$. 

For the interpolation between quantum and classical position and momentum observables we define the ``sharpened observables'' as
\ba\label{P1}
X_\beta&=&\cos^2 \beta X_Q+\sin^2\beta X_{cl}=X_{cl}+\frac12\cos^2\beta X_s,\nn\\
P_\beta&=&\cos^2\beta P_Q+\sin^2\beta P_{cl}=P_{cl}+\frac12\cos^2\beta P_s,
\ea
with ``sharpness parameter'' $s=\sin^2\beta$ between zero for quantum measurements and one for classical measurements. In consequence, also the commutation relation between $X_\beta$ and $P_\beta$ interpolates continuously between the quantum particle for $\beta=0$ and the classical particle for $\beta=\pi/2$, 
\be\label{P2}
[P_\beta,X_\beta]=-i\cos^2\beta.
\ee

For the time evolution of the interpolated observables we employ the Heisenberg picture of the quantum  formalism. Using the commutation relations \eqref{P3} and 
\ba\label{174A}
i[H_L,X_Q]=\frac1m P_Q~,~i[H_W,X_{cl}]=\frac1m P_{cl},\nn\\
i[H_L,P_Q]=-V'(X_{cl})-V''(X_{cl})(X_Q-X_{cl}),\nn\\
i[H_W,P_{cl}]=-\frac12\big (V'(X_Q)+V'(2X_{cl}-X_Q)\big).
\ea
allows us to compute the time evolution of the interpolated operators $X_\beta$ and $P_\beta$ in dependence on the choice of the Hamiltonian $H_L$ or $H_W$ or linear combinations thereof.If the correlation functions involving $X_\beta$ and $P_\beta$ are known for some initial time $t_0$, the time evolution in the Heisenberg picture determines these correlations for $t\neq t_0$. 

We will concentrate here on the special case of the harmonic oscillator for which the quantum and classical time evolution coincide. Indeed, for the special case of a harmonic potential, $V=(c/2)x^2$, one has 
\be\label{P3A}
H_W=H_L=\frac 1m P_{cl} P_s+c X_{cl}X_s.
\ee
Thus $X_\beta, P_\beta$ obey the standard time evolution according to 
\ba\label{P4}
\dot X_\beta&=&i[\tilde H,X_\beta]=\frac 1m P_\beta,\nn\\
\dot P_\beta&=&i[\tilde H,P_\beta]=-V'(X_\beta)=-c X_\beta.
\ea
The time evolution for the sharpened position observable is given in this case for all $\beta$ by
\be\label{P4A}
\kl\ddot X_\beta\kr=-\frac 1m\kl V'(X_\beta)\kr=-\frac cm\kl X_\beta\kr.
\ee
Also the time evolution of the dispersion is independent of $\beta$, as may be verified from the commutation relations
\ba\label{P4B}
i[H_W,X^2_\beta]&=&\frac 1m\{P_\beta,X_\beta\},\nn\\
i[H_W,P^2_\beta]&=&-c\{P_\beta,X_\beta\},\nn\\
i[H_W,\{P_\beta,X_\beta\}]&=& -2c X^2_\beta+\frac 2m P^2_\beta.
\ea
This extends to appropriately symmetrized higher correlation functions whose time evolution can be computed from the operator equations \eqref{P4} without use of the commutator \eqref{P2}. This shows that for free particles or a harmonic potential one cannot determine $\beta$ from the time evolution of symmetrized correlations for position and momentum. On this level we cannot distinguish quantum and classical observables. 

The issue is different if the probability distribution is known. Consider a harmonic oscillator in the quantum  ground state, with probability distribution $w_0$ given by eq. \eqref{P11} and $\kl X_\beta\kr=\kl P_\beta\kr=0$. We find for the dispersions
\ba\label{222}
\kl X^2_\beta\kr &=&\int_{z,p}\psi_0(z,p)(z+\frac i2\cos^2\beta\partial_p)^2
\psi_0(z,p)\nn\\
&=&\int_{z,p}\left[z^2+\cos^4\beta\left(\frac{1}{2m\omega}-\frac{p^2}{m^2\omega^2}\right)
\right]
w_0(z,p)\nn\\
&=&\frac{1}{4m\omega}(1+\cos^4\beta)
\ea
and similarly
\be\label{223}
\kl P^2_\beta\kr=\frac{m\omega}{4}
(1+\cos^4\beta).
\ee
We observe that the dispersion is minimal for the classical observables, $\cos^2\beta=0$. In particular, the uncertainty product 
\be\label{224}
(\kl X^2_\beta\kr\kl P^2_\beta\kr)^{1/2}=\frac{1+\cos^4\beta}{4}
\ee
admits for sharpened observables a smaller value than allowed by the quantum mechanical uncertainty relation, which is saturated for $\beta=0$, $\cos^4\beta=1$. One may therefore test experimentally if a given measurement realizes a pure quantum measurement or a sharpened measurement. For example, one may measure the position and/or momentum dispersion for a molecule in the ground state. For molecules which are well approximated by a harmonic potential one can extract from the relations \eqref{222}-\eqref{224} a bound for the sharpness parameter $\sin^2\beta$. We notice that the value of $\beta$ may depend on the choice of the device for position and momentum measurements. Different types of apparatus may realize different types of measurements. 

Beyond the special case of harmonic potentials the time evolution of expectation values and correlations involving $X_\beta$ and $P_\beta$ differs from the ones involving $X_Q$ and $P_Q$. This can be used for an experimental determination of $\beta$ or for putting bounds on $\beta$. 

\section{Zwitters - continuous interpolation of evolution equation}
\label{Zwitters - continuous}

Zwitters interpolate between the classical and quantum time evolution. We consider here a particular interpolation by using for the evolution equation of the classical wave function $i\partial_t\psi_C=H_\gamma\psi_C$, with interpolating Hamiltonian 
\ba\label{225}
H_\gamma&=&\frac{2}{m}P_{cl}(P_Q-P_{cl})+\cos^2\gamma\big(V(X_Q)-V(2X_{cl}-X_Q)\big)\nn\\
&&+2\sin^2\gamma V'(X_{cl})(X_Q-X_{cl})\nn\\
&=&\cos^2\gamma H_W+\sin^2\gamma H_L.
\ea
For $\gamma=0$ we recover the quantum time evolution $H_\gamma=H_W$, while $\gamma=\pi/2$ realizes the classical time evolution $H_\gamma=H_L$. For harmonic potentials $H_\gamma$ is independent of $\gamma$. Zwitters correspond to intermediate values of $\gamma$ in case of unharmonic potentials. 

Let us now concentrate on intermediate values of $\gamma$ where $\cos^2\gamma$ differs from $1$ and $0$. We may first ask if a conserved energy exists that commutes with the zwitter Hamiltonian $H_\gamma$ and equals the quantum energy $H_Q$ for $\gamma=0$ and the classical energy $H_{cl}$ for $\gamma=\pi/2$. We show in the appendix that this is not the case. The absence of a conserved energy is compatible with time translation symmetry, since the generator of the time translations is the trivially conserved $H_\gamma$. We will explore the consequences of this important property of zwitters below.

The choice of the interpolating Hamiltonian $H_\gamma$ is not unique. For example, one could replace the potential term in eq. \eqref{GP2} by $\cos^{-2}\gamma \left[V\left(x+\frac{i\cos^2\gamma}{2}\partial_p\right)-V\left(x-\frac{i\cos^2\gamma}{2}\partial_p \right)\right]$. It is well conceivable that for small values of $\gamma$ the different interpolations lead to qualitatively similar results (or even coincide in leading order in $\gamma$ after a possible multiplicative rescaling). This can only be settled by more extended investigations. We do not enter this discussion here and rather concentrate on one specific example for a zwitter particle, as given by eq. \eqref{225}. 

\medskip\noindent
{\bf 1. \quad Static zwitter states}

Next we turn to an investigation of possible static states for zwitters. By definition of the time evolution the static states are eigenvalues of $H_\gamma$ with eigenvalue zero,
\be\label{230}
H_\gamma\psi=0.
\ee
For a generic unharmonic potential these states are not eigenstates of a suitable energy operator with nonzero eigenvalues. Whatever reasonable energy observable we define, the measurement values will be distributed with a nonzero width for static states, rather than being sharp. Correspondingly, eigenvalues of a given energy operator will not be static with respect to the time evolution induced by $H_\gamma$.

This situation has interesting consequences for a possible experimental detection of zwitters. The ground state of an atom or molecule is expected to correspond to a static probability distribution. For zwitters, this state does no longer have a sharp energy, in contrast to quantum mechanics. Experimental bounds on the width of the ground state can therefore limit the ``zwitter parameter'' $\sin^2\gamma$. 

As an example, we may consider the quantum energy $H_Q$. We choose a discrete basis of pure quantum eigenstates, $H_Q\psi_{Q,n}=E_n\psi_{Q,n}$. The static ground state can be expanded in a corresponding basis $\big (z=(x+y)/2)$,
\be\label{97A}
\bar\psi_0(z,p)=\sum_{n,m}\lambda_{nm}\psi_{nm}(z,p)~,~H_\gamma\bar\psi_0=0,
\ee
where the classical basis functions
\be\label{97B}
\psi_{nm}(z,p)=\int d(x-y)[e^{-ip(x-y)}]\psi_n(x)\psi^*_m(y)
\ee
obey
\ba\label{235}
H_Q\psi_{nm}(z,p)=E_n\psi_{nm}(z,p).
\ea
We observe that $H_\gamma$ does not commute with $H_Q$,
\ba\label{98A}
&&[H_\gamma,H_Q]=\frac{\sin^2\gamma}{2m}
\left\{2i\left[V'\left(\frac{X_Q+\tilde X_Q}{2}\right)\right.\right.\nn\\
&&\left.-V'(X_Q)+\frac12 V'' \left(\frac{X_Q+\tilde X_Q}{2}\right)(X_Q-\tilde X_Q)\right]P_Q\nn\\
&& +V''\left(\frac{X_Q+\tilde X_Q}{2}\right)-V''(X_Q)\nn\\
&&+\frac14 V'''\left. \left(\frac{X_Q+\tilde X_Q}{2}\right)(X_Q-\tilde X_Q)\right\}.
\ea
It is instructive to consider a cubic correlation $\sim  dx^3/6$ to a harmonic potential $V(x)$ (cf. eq. \eqref{233}). The commutator is then proportional to $d$
\ba\label{236}
~[H_\gamma,H_Q]=-\frac{id\sin^2\gamma}{16m}\big\{(X_Q-\tilde X^2_Q)^2,P_Q\big\}+\dots.
\ea
In consequence, besides $\lambda_{00}$ the expansion coefficients $\lambda_{nm}$ are also nonvanishing $(\sim d\sin^2\gamma)$ for excited states of $H_Q$ with $E_n>E_0$. 

\medskip\noindent
{\bf 2. \quad Ground state broadening for zwitters}

We conclude that the quantum  energy has a nonzero width in the static ground state
\ba\label{237}
&&\kl (H_Q-\kl H_Q\kr)^2\kr_0=\int_{z,p}\bar\psi_0(H_Q-\kl H_Q\kr)^2\bar\psi_0\\
&&=\sum_{n,m}(E_n-\sum_{p,q}E_p|\lambda_{pq}|^2)^2|\lambda_{nm}|^2
\sim d^2\sin^4\gamma>0.\nn
\ea
We can rewrite this formula in terms of positive coefficients
\be\label{237A}
c_n=\sum_{m}|\lambda_{nm}|^2\geq 0~,~\sum_nc_n=1,
\ee
observing $c_{n>0}\sim d^2\sin^4\gamma$. The average value of $\kl H_Q\kr$ is larger than $E_0$,
\ba\label{237B}
\kl H_Q\kr_0&=&\sum_{n,m}E_n|\lambda_{nm}|^2=\sum_nc_nE_n,\nn\\
\kl H_Q\kr_0-E_0&=&\sum_nc_n(E_n-E_0)\sim d^2\sin^4\gamma.
\ea
This yields the general formula for the width of the ground state energy
\ba\label{102A}
\kl \Delta E^2\kr_0&=&\kl H^2_Q\kr_0-\kl H_Q\kr^2_0\nn\\
&=&\sum_nc_n(E_n-\kl H_Q\kr_0)^2.
\ea
The static ground is no longer the state with the lowest possible value of the quantum energy but also contains small admixtures of excited states. This can be used to put experimental bounds on $\gamma$. On the other hand, if a system starts with the lowest energy eigenstate of $H_Q$, this state will change in time since $H_\gamma\psi_{n=0,m}\neq 0$. The system will ``spontaneously populate'' excited states of $H_Q$. 

\medskip\noindent
{\bf 3. \quad Effective zwitter potential}

Let us assume that measurements or interactions with other systems only involve the quantum operators which can be constructed from $X_Q$ and $P_Q$. All expectation values and correlations can then be expressed in terms of the coarse grained quantum density matrix $\rho_Q(x,x')$ defined by eq. \eqref{218}. We may therefore project the time evolution of the classical wave function onto a corresponding time evolution of $\rho_Q(x,x')$. This cannot be expressed in terms of $\rho_Q$ alone, since the evolution equation
\ba\label{238}
i\partial_t\rho_Q(x,x')&=&\big (H_Q(x)-H_Q(x')\big)\rho_Q(x,x')\\
&&+\sin^2\gamma\int_y\big(\Delta(x,y)-\Delta(x',y)\big)\sigma(x,x'y)\nn
\ea
involves the diagonal elements of the ``classical density matrix'' in position space, $\tilde\rho_C(x,x',y,y')$, namely
\be\label{239}
\sigma(x,x',y)=\tilde\rho_C(x,x',y,y)~,~\rho_Q(x,x')=\int_y\sigma(x,x',y).
\ee
In eq. \eqref{238} we define $\Delta(x,y)$ as
\be\label{240}
\Delta(x,y)=V'\left(\frac{x+y}{2}\right)(x-y)-V(x)+V(y).
\ee
In turn, $\sigma$ obeys the evolution equation
\ba\label{241}
i\partial_t\sigma(x,x',y)&=&\Big\{H_Q(x)-H_Q(x')+\sin^2\gamma\big (\Delta(x,y)\nn\\
&&-\Delta(x',y)\big)\Big\}\sigma(x,x',y).
\ea

For a pure quantum state one has the factorization property $\tilde\rho_C(x,x',y,y')=\rho_Q(x,x')\rho_Q(y',y)$ and this motivates the ansatz
\be\label{242}
\sigma(x,x',y)=\rho_Q(x,x')\rho_Q(y,y)+\delta(x,x',y).
\ee
Neglecting first $\delta$ one finds for the time evolution of the quantum density matrix a modified equation
\be\label{243}
i\partial_t\rho_Q=[H^{(\gamma)}_Q,\rho_Q],
\ee
with
\ba\label{244}
&&H^{(\gamma)}=H_Q+W_\gamma,\nn\\
&&W_\gamma=\sin^2\gamma \int_y\Delta(x,y)\rho_Q(y,y).
\ea
The time evolution remains unitary since $H^{(\gamma)}_Q$ is a hermitean operator. It is, however, no longer linear since the correction term $W_\gamma$ involves $\rho_Q$. Zwitters feel a modified effective potential
\be\label{245}
V_\gamma(x)=V(x)+W_\gamma(x).
\ee
We will call $V_\gamma(x)$ the ``zwitter potential'' since this is the potential that zwitter particles ``feel'' effectively in their time evolution according to eq. \eqref{243}. We recall that the zwitter potential is a dynamical quantity which depends on the state of the zwitter particle. The non-linearity of the effective evolution equation may be somewhat reminiscent of non-linear evolution equations for quantum condensates.

Let us consider a general quartic potential 
\be\label{233}
V(x)=a+bx+\frac c2 x^2+\frac d6 x^3+\frac{e}{24}x^4.
\ee
For example, this may describe the first terms of a Taylor expansion for small deviations from a harmonic potential. The ``dynamical term'' in the zwitter potential reads
\ba\label{246}
&&W_\gamma(x)=\frac{\sin^2\gamma}{24}
\big\{(3\kl X_Q\kr x^2-3\kl X^2_Q\kr x-x^3+\kl X^3_Q\kr)d\nn\\
&&+(\kl X_Q\kr x^3-\kl X^3_Q\kr x-\frac12 x^4+\frac12\kl X^4_Q\kr)e\big\}.
\ea
This shows the general feature that $V_\gamma$ does not only depend on $x$, but also on the expectation values $\kl X^n_Q\kr$ of various powers of $X_Q$. This reflects the nonlinear character of the time evolution since the expectation values $\kl X^n_Q\kr$ depend in turn on $\rho_Q(y,y)$ or, for a pure quantum state, on $|\psi_Q(y)|^2$. 

\medskip\noindent
{\bf 4. \quad Approximate zwitter ground state}

For small $\gamma$ we can compute iteratively the static solution which corresponds to the lowest value of $\kl H_Q\kr$. This state will be identified with the zwitter ground state. For this purpose we start with the ground state of the quantum  Hamiltonian $H_Q$, corresponding to $\gamma=0$. Next we evaluate $\kl X^n_Q\kr$ in the ground state of $H_Q$. This yields the lowest order approximation for the dynamic term in the zwitter potential $W_\gamma$. For the first correction to the zwitter ground state we construct the solution $\psi^{(\gamma)}_0$ of the stationary Schr\"odinger equation with the new potential $V_\gamma$. This corresponds to a static density matrix $\rho^{(\gamma)}_{Q,0}$, classical wave function $\psi^{(\gamma)}_0(z,p)$ according to eq. \eqref{P9}, and classical probability distribution $w^{(\gamma)}_0=(\psi^{(\gamma)}_0)^2$. For small $\gamma$ this static state will be a very good approximation for the zwitter ground state. We observe that this static solution is an eigenstate of $H^{(\gamma)}_Q=H_Q+W^{(0)}_\gamma$ with energy $E^{(\gamma)}_0$.

For the example of a quartic potential $(d=0)$ the quartic coefficient in the zwitter potential $V^{(0)}_\gamma$ is shifted by a factor $1-\sin^2\gamma/2$ as compared to $V$ and there is a constant shift $\sim \kl X^4_Q\kr$. The correction to the energy in lowest order
\be\label{112}
E^{(\gamma)}_0-E_0=\kl W_\gamma\kr
\ee
vanishes since $\kl W_\gamma\kr=0$. In lowest order perturbation theory one therefore has
\be\label{248}
\kl H_Q\kr=E^{(\gamma)}_0=E_0,
\ee
such that the mean energy is changed at most by effects $\sim \sin^4\gamma$, in accordance with eq. \eqref{237B}. Furthermore, $H_Q$ will not have a sharp value in the static state, since $\psi^{(\gamma)}_0$ is not an eigenstate of $H_Q$. 

\medskip\noindent
{\bf 5. \quad Zwitter ground state for Coulomb potential}

It is clear that strong experimental bounds on $\gamma$ require that the potential $V(x)$ is known with high precision in addition to the feasibility of high precision energy measurements. A good candidate is the Coulomb-potential $V(\vec r)=c/|r|$, with $c=-e^2/4\pi=-\alpha$ for the hydrogen atom. It leads to a correction \eqref{238} of the time evolution equation with
\be\label{249}
\Delta(\vec x,\vec y)=-c
\left(4\frac{\vec x^2-\vec y^2}{|\vec x+\vec y |^3}+\frac{1}{|\vec x|}-\frac{1}{|\vec y|}\right).
\ee
In order to compute $W^{(0)}_\gamma$ we have to evaluate $\kl \Delta\kr_0$ in the ground state $\psi_{Q,0}$ of $H_Q$,
\be\label{250}
W^{(0}_\gamma(\vec  x)=\sin^2\gamma \int_y\psi^*_{Q,0}(\vec y)\Delta(\vec x,\vec y)\psi_{Q,0}(\vec y).
\ee
Since $\psi_{Q,0}$ depends only on $|\vec y|$, the only $\vec x$-dependence of the angular integration in eq. \eqref{250} involves
\ba\label{251}
&&\hspace{-1.0cm}\int_y|\vec x+\vec y|^{-3}f(|y|)=2\pi\int^\infty_0d|y|f(|y|)\nn\\
&&\int^1_{-1}d\cos\varphi~ (|x|^2+|y|^2+2|x||y|\cos\varphi)^{-3/2}\\
&&=2\pi\int^\infty_0\frac{d|y|f|y|)}{|x||y|}
\left(\frac{1}{||x|-|y||}-\frac{1}{|x|+|y|}\right).\nn
\ea
Neglecting $x$-independent terms this yields our result for the zwitter potential
\ba\label{252}
W^{(0)}_\gamma&=&-c\sin^2\gamma \frac{f(|x|)}{|x|},\\
f(|x|)&=&5-16\pi\int^\infty_{|x|}d|y||\psi_{Q,0}(|y|)|^2\left(1+\frac{|x|}{|y|}\right),\nn
\ea
and we observe
\be\label{252}
f(|x|\to 0)=1~,~f(|x|)\to \infty)=5.
\ee
Within reasonable accuracy we can replace $f(x)$ by a constant $\bar f$ within the interval $1\leq \bar f\leq 5$. 

In this approximation the static state which corresponds to the lowest eigenvalue of $H_Q+W^{(0)}_\gamma$ corresponds to the usual ground state wave function, but with a shifted constant in the Coulomb potential 
\be\label{118A}
V_\gamma(\vec x)=\frac{\tilde c}{|x|}~,~\tilde c=c(1-\sin^2\gamma\bar f).
\ee
For the hydrogen atom, the shifted Bohr radius and shifted Rydberg constant result in a zwitter ground state wave function
\be\label{118B}
\psi^{(\gamma)}_0(x)=
(\pi a^{(\gamma)3})^{-\frac12}
\exp \left\{-\frac{|x|}{a^{(\gamma)}}\right\},
\ee
where
\be\label{254}
a^{(\gamma)}=\frac{a}{1-\sin^2\gamma\bar f},
\ee
and $a=\alpha/\mu$ is the Bohr radius. 

\medskip\noindent
{\bf 6.\quad Ground state broadening for Coulomb 

\hspace{0.4cm}potential}

One of the most interesting quantities for experimental tests of $\gamma$ may be the energy width of the zwitter ground state $\Delta E$, since for $\gamma=0$ the quantum ground state has a sharp energy $(\Delta E=0)$. We compute $\Delta E$ for the Coulomb potential in the approximation where the zwitter ground state obeys eq. \eqref{254}. For this purpose we expand the zwitter ground state \eqref{254} in terms of the usual eigenstates of the quantum observables $H_Q,L^2$ and $L_z$,
\be\label{255}
\psi^{(\gamma)}_0(x)=\sum_{n,l,m}a_{nlm}\psi_{nlm}(x),
\ee
with 
\ba\label{261A}
a_{100}&=&\int_x\psi^*_{100}(x)\psi^{(\gamma)}_0(x)\nn\\
&=&4(a^{(\gamma)}a)^{-\frac32}\int^\infty_0 drr^2\exp 
\left\{-\left(\frac 1a+\frac{1}{a^{(\gamma)}}\right)r\right\}\nn\\
&=&\frac{8(a^{(\gamma)}a)^{3/2}}{\big (a^{(\gamma)}
+a\big )^3}\approx 1-\frac 38\sin^4\gamma\bar f^2,
\ea
and
\ba\label{256}
a_{200}&=&\int_x\psi^*_{200}(x)\psi_0^{(\gamma)}(x)\nn\\
&=&\sqrt{2}(a^{(\gamma)}a)^{-3/2}\int^\infty_0 dr r^2
\left(1-\frac{r}{2a}\right)\nn\\
&&\exp \left\{-\left(\frac{1}{2a}+\frac{1}{a^{(\gamma)}}\right)r\right\}\nn\\
&=&2\sqrt{2}(a^{(\gamma)}a)^{3/2}(a-a^{(\gamma)})
\left(a+\frac12 a^{(\gamma)}\right)^{-4}\nn\\
&&\approx-\frac{32\sqrt{2}}{81}\sin^2\gamma\bar f,\nn\\
a_{21m}&=&0.
\ea
More generally, angular momentum conservation implies $a_{nlm}=0$ for $l\neq 0$. The static state $\psi^{(\gamma)}_0$ is a superposition of $s$-wave states. The coefficients $c_n$ in eq. \eqref{237B} are given by $c_n=|a_{n00}|^2$. 

We can compute the average of the quantum energy in the static state \eqref{254} directly $(c=-e^2/4\pi=-\alpha$ for the hydrogen atom)
\ba\label{257}
\kl H_Q\kr&=&\int_x\psi^{(\gamma)}(x)
\left(-\frac{\Delta}{2\mu}+\frac{c}{|x|}\right)\psi^{(\gamma)}_0(x)\nn\\
&=&-\frac{c^2\mu}{2}(1-\sin^2\gamma\bar f)^2+c\sin^2\gamma \bar f\kl \frac{1}{|x|}\kr\nn\\
&=&-\frac{c^2\mu}{2}(1-\sin^4\gamma\bar f^2)=E_0+\delta E^{(\gamma)},
\ea
with $E_0$ the ground state energy of the quantum hydrogen atom $(\gamma=0)$. In order to derive eq. \eqref{257} we use $\kl 1/|x|\kr=1/a^{(\gamma)}=-(1-\sin^2\gamma\bar f)c\mu$. We observe the relative shift in the mean energy
\be\label{258}
\frac{\delta E^{(\gamma)}}{|E_0|}=\sin^4\gamma\bar f^2.
\ee
The energy width of the static state obtains as 
\ba\label{259}
\Delta E&=&(\kl H^2_Q\kr-\kl H_Q\kr^2)^{1/2}\nn\\
&=&|c|\sin^2\gamma\bar f
\left(\kl\frac{1}{|x|^2}\kr-
\kl\frac{1}{|x|}\kr^2\right)^{1/2}\nn\\
&\approx&c^2\mu \sin^2\gamma\bar f=2\sin^2\gamma \bar f|E_0|.
\ea
The relative energy broadening is proportional to $\sin^2\gamma$ 
\be\label{125A}
\frac{\Delta E}{|E_0|}=2\sin^2\gamma\bar f.
\ee
For small $\sin^2\gamma$ the energy shift $\delta E^{(\gamma)}$ is much smaller than the energy width $\Delta E$ of the static state. It is suppressed by a further factor $\sin^2\gamma$,
\ba\label{260}
\frac{\delta E^{(\gamma)}}{\Delta E}=\frac12\sin^2\gamma\bar f.
\ea

\newpage\noindent
{\bf 7. \quad Coarse grained static state}

In lowest order in $\sin^2\gamma$ the association of the zwitter ground state with the static state with lowest $\kl H_Q\kr$ seems rather appealing. Going beyond this order requires more thought what one means by a zwitter ground state. In particular, it is not obvious if the concept of a ``ground state'' for zwitters should be given by a completely static probability distribution. We rather pursue here the notion of a ``coarse grained static state'', for which only the $y$-integral of the classical density matrix is time-independent
\ba\label{261}
\partial_t\int_y\tilde \rho_C(x,x',y,y)=0.
\ea
We may call this a quasi-static state or coarse grained static state. Obviously, the condition \eqref{261} is much weaker than the one for a static probability distribution, i.e. $\partial_t\rho(x,x',y,y')=0$. The condition \eqref{261} is sufficient for static expectation values of all quantum observables, since it corresponds to a static quantum density matrix, $\partial_t\rho_Q(x,x')=0$. In other words, a coarse grained static state allows for time-fluctuations of the classical probability distribution, provided that the coarse grained density matrix $\rho_Q(x,x')$ remains static.

Using the Hamiltonian $H_\gamma$ for the time evolution of zwitters the condition \eqref{261} reads
\ba\label{262}
&&\int_y\Big\{ H_Q(x)-H_Q(x')+\sin^2\gamma\big (\Delta(x,y)-\Delta(x',y)\big)\Big\}\nn\\
&&\hspace{4.0cm}\times\tilde\rho_C(x,x',y,y)=0.
\ea
Inserting eq. \eqref{242} one obtains the equivalent condition
\ba\label{263}
\big (H^{(\gamma)}_Q(x)-H^{(\gamma)}_Q(x')\big)\rho_Q(x,x')=-\sin^2\gamma Q(x,x'),
\ea
with
\ba\label{264}
Q(x,x')=\int_y\big(\Delta(x,y)-\Delta(x',y)\big)\delta(x,x',y).
\ea
In turn, the time evolution of $Q$ obeys
\ba\label{265}
i\partial_t Q(x,x')&=&\big (H^{(\gamma)}_Q(x)-H^{(\gamma)}_Q(x')\big)
Q(x,x')\nn\\
&&+\sin^2\gamma R(x,x'),
\ea
with
\ba\label{283A}
&&\hspace{-1.0cm}R(x,x')=\int_y\rho(x,x',y,y)\Big[\big(\Delta(x,y)-\Delta(x',y)\big)^2\\
&&-\ll\Delta(x,y)-\Delta(x',y)\gg\big(\Delta(x,y)-\Delta(x',y)\big)\Big],\nn
\ea
and
\ba\label{266}
\ll\Delta(x,y)\gg~=\int_y\Delta(x,y)\rho_Q(y,y).
\ea

In general, $\partial_t Q$ does not vanish for $Q=0$ due to $R\neq 0$. Therefore the r.h.s. of eq. \eqref{263} cannot vanish for all $t$. However, we observe that $H^{(\gamma)}_Q$ is time independent if $\rho_Q$ is static. Thus the condition \eqref{263} requires for a coarse grained static state that $Q$ is time independent,  $\partial_t Q=0$. In turn, we conclude from the vanishing of the l.h.s. of eq. \eqref{265} that $Q\sim \sin^2\gamma$. For small $\sin^2\gamma$ we may therefore neglect the term $\sim \sin^2\gamma Q$ in eq. \eqref{263}. It amounts to a correction $\sim\sin^4\gamma$, while the leading difference between $H^{(\gamma)}_Q$ and $H_Q$ is of the order $\sin^2\gamma$ due to $W_\gamma$ \eqref{246}. In this approximation $\rho_Q$ corresponds to a static density matrix with respect to a unitary, but non-linear, quantum time evolution given by the modified Hamiltonian $H^{(\gamma)}_Q$ in eq. \eqref{244}. It seems likely that the states with lowest $\kl H_Q\kr$ correspond to the pure state discussed above. This discussion justifies the neglection of $\delta$ in eq. \eqref{242}. 

Experimental efforts for establishing a bound on the zwitter parameter $z=\sin^2\gamma$ may concentrate on the efforts of measuring or restricting a nonzero width of the ``zwitter ground state'' which corresponds to the coarse grained static state with lowest $\kl H_Q\kr$. Beyond the Coulomb potential, this may be particularly interesting for potentials for which the quantum particle has an almost degenerate ground state. One expects sizable effects for values of $\sin^2\gamma$ for which $\Delta E$ becomes comparable to the energy difference between the two lowest energies of the quantum particle, $E_1-E_0$. On the other hand, since $\sin^2\gamma$ is proportional to $1/E_0$ according to eq. \eqref{125A}, small $\Delta E$ combined with larger $E_0$ may give the most stringest bounds. In this context we note the long relaxation time of nuclear spin polarized $^3He$ which has been reported in ref. \cite{Gem}. This places a bound on the energy width of the ground state $\Delta E<0.6\cdot 10^{-20}$eV. If we take for $E_0$ the nuclear binding energy this would yield a bound $|\gamma|\lesssim 3\cdot 10^{-14}$, where we assume $\bar f$ in eq. \eqref{125A} to be of the order one. One may speculate that a minimal value for $\gamma^2$ could be associated with the ratio $(E_0/M_F)^n$, where $M_F$ is some scale of fundamental physics. If one identifies $M_F$ with the Planck mass the above bound for $\gamma$ would imply $n\geq 2$.

\section{Conclusions and discussion}
\label{conclusionsand}
In this paper we have described quantum particles in terms of classical probabilities in phase-space. This demonstrates that quantum physics can be formulated in terms of the basic concepts of a classical probabilistic theory. Quantum and classical particles can both be described within the same conceptual setting and mathematical formalism. Two basic features distinguish the dynamics and measurements of a quantum particle from a classical particle.

(i) The time evolution of the probability density in phase space is governed by a new basic dynamical law that replaces the Liouville equation. The fundamental time evolution equation remains a linear first order differential equation for the classical wave function $\psi_C(x,p)$, which is related to the probability density by $w(x,p)=\psi^2_C(x,p)$. In general, this equation involves higher orders in the momentum derivatives, however. It becomes therefore a non-linear equation when expressed in terms of the probability density $w$. In contrast to the Liouville equation, the ``quantum evolution'' of $\psi_C$ or $w$ is no longer compatible with the notion of sharp trajectories of particles. The classical trajectories and Newtons' laws emerge only in the classical limit when the action is large as compared to $\hbar$. This feature is familiar from quantum mechanics.

(ii)Measurements of properties of a quantum particle are related to quantum observables for position and momentum which differ from the classical observables. While the classical observables have a well defined value for every point $(x,p)$ in phase-space, this does not hold for the quantum observables. The quantum observables partly involve genuinely statistical properties of the probability distribution, as its roughness in position and momentum. These statistical aspects cannot be associated to a single point in phase space, and the quantum observables are therefore not simply functions of $x$ and $p$. On the level of the classical wave function $\psi_C(x,p)$ the quantum observables are associated to linear operators, while the expression in terms of $w (x,p)$ becomes non-linear.

Both "`quantum features"' (i) and (ii) are singled out by their consistence with a coarse graining of the classical probabilities, whereby part of the information contained in $w (x,p)$ is typically discarded. The quantum evolution and the quantum observables remain meaningful for the coarse grained subsystem of the statistical ensemble. This does not hold for the classical time evolution according to the Liouville equation, nor for the classical observables. We can therefore associate the characteristic quantum effects with the coarse graining to a subsystem.

On the level of the probability distribution in phase space $w (x,p)$ or the associated classical wave function $\psi_C(x,p)$ the classical observables are well defined for quantum particles, and the quantum observables are well defined for classical particles. This raises the question which observables can be associated to which type of measurement. We have investigated the possibility of sharpened observables that interpolate continuously between the quantum observables and classical observables. For the sharpened observables Heisenberg's uncertainly relation is weakened.

The common conceptual and formal setting for quantum particles and classical particles also allows for a continuous interpolation of the basic dynamical law between the classical evolution and the quantum evolution. Zwitters are particles for which the basic evolution law is neither purely classical nor purely quantum, but rather in-between. We have characterized this interpolation by a zwitter parameter $\sin^2\gamma$ which interpolates between the quantum evolution for $\gamma=0$ and the classical evolution for $\gamma=\pi/2$. For all values of $\gamma$ we have the same conceptual setting based on the probability distribution in phase-space. This differs from other, more formal, interpolations, as replacing $\hbar$ by $\hbar\cos^2\gamma$ in the time evolution of the Wigner function.

Zwitters are interesting candidates for an effective one-particle description of classical or quantum collective states as droplets of a liquid or a Bose-Einstein condensate. The characterization by a single particle means that the time evolution can be described by a probability density or classical wave function in phase space, without the need of adding further information. We believe that for a wide range of circumstances the deviation of the dynamics of such objects from pointlike classical particles or from quantum particles can be cast into the formalism for zwitters. We hope that zwitters may yield an adequate and rather simple description of experiments and physical situations involving such collective states. This may cover collective classical states as droplets of a liquid or dust particles that do not follow precisely the classical trajectories of point particles. Zwitters may also account for collective quantum states as a Bose-Einstein condensate or macromolecules. For these effective one-particle descriptions zwitters are not fundamental, and the value of $\hbar$ can reflect some macroscopic property.

On the other hand, zwitters may also be used for fundamental tests of quantum mechanics. For small values of $\gamma$ consistent theories very close to quantum mechanics can be realized. This permits precision tests for the validity of quantum physics by putting experimental bounds on the zwitter parameter $z=\sin^2\gamma$. Potential candidates for obtaining strong bounds on $z$ are precision measurements of the energy-width of the ground state of atoms, molecules,  or other quantum objects. We have discussed this issue in detail. Other venues may be precision tests of quantum interference experiments or tunneling. We hope that our approach opens the door for interesting experimental tests how well quantum mechanics works quantitatively, rather than staying with yes/no decisions between quantum physics and classical physics.

\LARGE
\section*{APPENDIX A: ABSENCE OF CONSERVED ENERGY FOR ZWITTERS}
\renewcommand{\theequation}{A.\arabic{equation}}
\setcounter{equation}{0}

\normalsize
In this appendix we argue that for zwitters defined by the particular interpolating Hamiltonian $H_\gamma$ \eqref{225} no conserved energy exists (besides the trivially conserved generator of time translations $H_\gamma$ itself). This involves the search for an energy operator $\hat H$ which commutes with $H_\gamma$. We have therefore to explore the commutation relations for different pieces that could be additive building blocks of $\hat H$ with $H_W$ and $H_L$. 

With 
\ba\label{226}
H_L&=&H_{kin}+V_L~,~H_{kin}=\frac{1}{2m}(P^2_Q-\tilde P^2_Q),\nn\\
V_L&=&V'\left(\frac{X_Q+\tilde X_Q}{2}\right)(X_Q-\tilde X_Q)\nn\\
H_W&=&H_Q-\tilde H_Q,
\ea
one has the commutation relations
\ba\label{229A}
~[H_Q,\tilde H_Q]&=&0~,~[H_{cl},H_L]=0,\nn\\
~[H_Q,H_L]&=&\frac{1}{2m}[P^2_Q,V_L]-[H_{kin},V(X_Q)],\nn\\
~[\tilde H_Q,H_L]&=&\frac{1}{2m}[\tilde P^2_Q,V_L]-[H_{kin},V(\tilde X_Q)],
\ea
and
\ba\label{227}
~[H_Q,H_{cl}]&=&\frac{1}{2m}\Big[P^2_Q,V\left(\frac{X_Q+\tilde X_Q}{2}\right)\Big]\nn\\
&&-\frac{1}{8m}\left[\left(P^2_Q+\tilde P^2_Q-2P_Q\tilde P_Q\right),V(X_Q)\right],\nn\\
~[\tilde H_Q,H_{cl}]&=&\frac{1}{2m}\Big[\tilde P^2_Q,V
\left(\frac{X_Q+\tilde X_Q}{2}\right)\Big]\\
&&-\frac{1}{8m}\Big [(P^2_Q+\tilde P^2_Q-2P_Q\tilde P_Q),V(\tilde X_Q)\Big ].\nn
\ea
We notice that the last four commutators typically differ from zero such that the existence of a conserved energy is not obvious. 

Next we investigate the commutation relations for different possible pieces composing $\hat H$. We list
\ba\label{230A}
&&[P^2_Q,V(X_Q)]=-2iV'(X_Q)P_Q-V''(X_Q),\nn\\
&&[P_Q\tilde P_Q~,~V(X_Q)]=-iV'(X_Q)\tilde P_Q,
\ea
and
\ba\label{230B}
&&\left [P^2_Q,V\left (\frac{X_Q+\tilde X_Q}{2}\right)\right]=\nn\\
&&-iV'\left(\frac{X_Q+\tilde X_Q}{2}\right)P_Q-\frac14 V''
\left(\frac{X_Q+\tilde X_Q}{2}\right),\nn\\
&&\left [P_Q\tilde P_Q~,~V\left(\frac{X_Q+\tilde X_Q}{2}\right)\right]=-\frac i2 V'
\left(\frac{X_Q+\tilde X_Q}{2}\right)\nn\\
&&\hspace{1.9cm}(P_Q+\tilde P_Q)-\frac14 V''\left(\frac{X_Q+\tilde X_Q}{2}\right),
\ea
as well as
\ba\label{230C}
&&\hspace{-1.3cm}[P^2_Q,V_L]=-2iV'\left(\frac{X_Q+\tilde X_Q}{2}\right)P_Q\nn\\
&&-V''\left(\frac{X_Q+\tilde X_Q}{2}\right)\big (1+i(X_Q-\tilde X_Q)P_Q\big)\nn\\
&&-\frac14V'''\left(\frac{X_Q+\tilde X_Q}{2}\right)(X_Q-\tilde X_Q),\nn\\
&&\hspace{-1.3cm}[\tilde P^2_Q,V_L]=2iV'\left(\frac{X_Q+\tilde X_Q}{2}\right)\tilde P_Q\nn\\
&&+V''\left(\frac{X_Q+\tilde X_Q}{2}\right)\big(1-i(X_Q-\tilde X_Q)\tilde P_Q\big)\nn\\
&&-\frac14 V'''\left(\frac{X_Q+\tilde X_Q}{2}\right)(X_Q-\tilde X_Q),
\ea
and
\ba\label{228}
&&\hspace{-1.3cm}[P_Q\tilde P_Q,V_L]=iV'\left(\frac{X_Q+\tilde X_Q}{2}\right)(P_Q-\tilde P_Q)\nn\\
&&-\frac i2 V''\left(\frac{X_Q+\tilde X_Q}{2}\right)
(X_Q-\tilde X_Q)(P_Q+\tilde P_Q)\nn\\
&&-\frac14 V'''\left(\frac{X_Q+\tilde X_Q}{2}\right)(X_Q-\tilde X_Q),
\ea

For the energy observable $\hat H$ we make the ansatz of a linear combination
\ba\label{228A}
\hat H&=&\frac{A}{2m}P^2_Q+\frac{\tilde A}{2m}\tilde P^2_Q+\frac{B}{m}P_Q\tilde P_Q+EV(X_Q)\nn\\
&&+\tilde EV(\tilde X_Q)+FV\left(\frac{X_Q+\tilde X_Q)}{2}\right).
\ea
This yields the commutation relation with $H_\gamma$
\ba\label{229}
&&\hspace{-0.5cm}[H_\gamma,\hat H]=\frac{1}{2m}\Big\{2i(A\cos^2\gamma-E)V'(X_Q)P_Q\nn\\
&&-2i(\tilde A\cos^2\gamma
-\tilde E)V'(\tilde X_Q)\tilde P_Q\nn\\
&&+2iB\cos^2\gamma\big(V'(X_Q)\tilde P_Q -V'(\tilde X_Q)P_Q\big)\nn\\
&&+(A\cos^2\gamma-E)V''(X_Q)-(\tilde A\cos^2\gamma-\tilde E)V''(\tilde X_Q)\nn\\
&&+[2i(A-B)\sin^2\gamma-iF]V'\left(\frac{X_Q+\tilde X_Q}{2}\right) P_Q\nn\\
&&-\big [2i(\tilde A-B)\sin^2\gamma-iF\big]V'\left(\frac{X_Q+\tilde X_Q}{2}\right)
\tilde P_Q\nn\\
&&+(A-\tilde A)\sin^2\gamma V''\left(\frac{X_Q+\tilde X_Q}{2}\right)\\
&&+i\sin^2\gamma V''\left(\frac{X_Q+\tilde X_Q}{2}\right)
(X_Q-\tilde X_Q)\nn\\
&&\hspace{1.7cm}[(A+B)P_Q+(\tilde A+B)\tilde P_Q]\nn\\
&&+\frac14(A+\tilde A+2B)\sin^2\gamma V'''
\left(\frac{X_Q+\tilde X_Q}{2}\right)(X_Q-\tilde X_Q)\Big\}.\nn
\ea
The classical Hamiltonian corresponds to $A=\tilde A=1/4,~B=-1/4,~E=\tilde E=0,~F=1$ and commutes with $H_\gamma$ for $\sin^2\gamma=1$, while the quantum Hamiltonian obeys $A=E=1,~\tilde A=B=\tilde E=F=0$, and commutes with $H_\gamma$ for $\cos^2\gamma=1$. For intermediate values of $\gamma$ and generic unharmonic potentials the relation $[H_\gamma,\hat H]=0$ has no nontrivial solution. Adding further terms to $\hat H$ does not change the situation. Generically, no nontrivial conserved energy exists for zwitters of the type defined by eq. \eqref{225}. 

As an example we may consider the ``zwitter energy''
\be\label{231}
E_\gamma=\cos^2\gamma H_Q+\sin^2\gamma H_{cl},
\ee
which interpolates between the quantum energy for $\gamma=0$ and the classical energy for $\gamma=\pi/2$. This corresponds to $A=\cos^2\gamma+\sin^2\gamma/4,\tilde A=-B=\sin^2\gamma/4,~E=\cos^2\gamma,\tilde E=0,~F=\sin^2\gamma$ and we infer the commutator
\ba\label{232}
~[H_\gamma,E_\gamma]&=&\frac{\cos^2\gamma\sin^2\gamma}{2m}
\Bigg\{i\left[ V'\left(\frac{X_Q+\tilde X_Q}{2}\right)\right.\nn\\
&&\left. -\frac12 V'(X_Q)-\frac12 V'(\tilde X_Q)\right](P_Q+\tilde P_Q)\nn\\
&&+i[V'(\tilde X_Q)-V'(X_Q)]P_Q\nn\\
&&+V''\left(\frac{X_Q+\tilde X_Q}{2}\right)-\frac34 V''(X_Q)-\frac14 V''(\tilde V_Q)\nn\\
&&+iV''\left(\frac{X_Q+\tilde X_Q}{2}\right)(X_Q-\tilde X_Q)P_Q\nn\\
&&+\frac14 V'''\left(\frac{X_Q+\tilde X_Q}{2}\right)(X_Q-\tilde X_Q)\Bigg\}.
\ea
Performing a Taylor expansion of $V$ using eq. \eqref{233} we find that in leading order $(e=0)$ only the cubic term contributes to the commutator
\ba\label{234}
~[H_\gamma,E_\gamma]=-\frac{id\cos^2\gamma\sin^2\gamma}{16m}
(X_Q-\tilde X_Q)^2(P_Q+\tilde P_Q).\nn\\
\ea
As it should be, the commutator vanishes for $\gamma=0,\pi/2$ and for a harmonic potential.


\end{document}